\documentclass{article}

\usepackage{cite}

\begin{document}


\newcounter{zeqn}
\setcounter{zeqn}{1}

\newcommand{\zeqnalt}{\renewcommand{\theequation}
{\arabic{section}.\arabic{equation}{\rm\alph{zeqn}}}}
\newcommand{\zeqnapp}{\renewcommand{\theequation}
{{\rm A}.\arabic{equation}}}
\newcommand{\zeqnorg}{\renewcommand{\theequation}
{\arabic{section}.\arabic{equation}}}
\newcommand{\zadeqn}{\addtocounter{zeqn}{1}\addtocounter{equation}{-1}}
\newcommand{\zsetc}{\setcounter{zeqn}{1}}
\newcommand{\zeqnAalt}{\renewcommand{\theequation}
{{\rm A}.\arabic{equation}{\rm\alph{zeqn}}}}





\newcommand{\ignorethis}[1]{}
\newcommand{\zbe}{\begin{equation}}
\newcommand{\zee}{\end{equation}}


\newcommand{\osp}{{\it{osp}}$(2|2)$}
\newcommand{\ospa}{{\it{osp}}$(2|2)^{(1)}$}
\newcommand{\uosp}{${\it U_q[}$\osp $]$}
\newcommand{\uospa}{${\it U_q[}$\ospa $]$}
\newcommand{\uga}{${\it U_q(}\hat{g}{\it )}$}
\newcommand{\zs}{{\it{S}}}
\newcommand{\ztS}{\tilde{S}}
\newcommand{\zr}{{\it R}}
\newcommand{\zss}{${\it S}$}
\newcommand{\zrr}{${\it R}$}
\newcommand{\rt}{\check{R}}

\newcommand{\scft}{S_{\rm cft}}

\newcommand{\pp}{\psi_+}
\newcommand{\psm}{\psi_-}
\newcommand{\pbp}{\overline{\psi}_+}
\newcommand{\pbm}{\overline{\psi}_-}
\newcommand{\zpl}{\psi_\pm}
\newcommand{\zpr}{\overline{\psi}_\pm}
\newcommand{\pz}{\Psi}
\newcommand{\pzb}{\overline{\Psi}}
\newcommand{\pzh}{\hat{\Psi}}
\newcommand{\bt}{\beta}
\newcommand{\btb}{\overline{\beta}}
\newcommand{\ga}{\gamma}
\newcommand{\gab}{\overline{\gamma}}
\newcommand{\gah}{\hat{\gamma}}

\newcommand{\zd}{\partial}
\newcommand{\zdd}{\bar{\partial}}
\newcommand{\zb}{{\overline{z}}}
\newcommand{\wb}{{\overline{w}}}
\newcommand{\zm}{\frac{d^2 x}{2\pi}}
\newcommand{\zmw}{\frac{d^2 w}{2\pi i}}
\newcommand{\zmwb}{\frac{d^2 \overline{w}}{2\pi i}}
\newcommand{\zdz}{\partial_z}
\newcommand{\zdb}{\partial_{\overline{z}}}

\newcommand{\fa}{\frac{1}{2}}

\newcommand{\gh}{\widehat{G}}
\newcommand{\kh}{\widehat{K}}
\newcommand{\gb}{\overline{G}}
\newcommand{\ghb}{\overline{\gh}}
\newcommand{\kb}{\overline{K}}
\newcommand{\khb}{\overline{\kh}}
\newcommand{\jb}{\overline{J}}
\newcommand{\hb}{\overline{H}}
\newcommand{\qb}{\overline{Q}}
\newcommand{\jbz}[1]{{\overline{J}}^{{}_{\scriptstyle #1}}}
\newcommand{\qbz}[1]{{\overline{Q}}^{{}_{\scriptstyle #1}}}
\newcommand{\hbz}[1]{{\overline{H}}^{{}_{\scriptstyle #1}}}

\newcommand{\al}{\alpha}
\newcommand{\bh}{\hat{\beta}}
\newcommand{\zo}{{\cal O}}
\newcommand{\zq}{{\cal Q}}

\newcommand{\zpp}{\Phi_{\rm pert.}}
\newcommand{\zps}{\phi_{\rm pert.}}
\newcommand{\zpsb}{\overline{\phi}_{\rm pert.}}
\newcommand{\zp}{\phi}
\newcommand{\zpb}{\overline{\phi}}
\newcommand{\zpv}{\vec{\phi}}
\newcommand{\zpbv}{\vec{\overline{\phi}}}
\newcommand{\zPv}{\vec{\Phi}}
\newcommand{\zPtv}{\vec{\tilde{\Phi}}}

\newcommand{\zqb}[1]{{\overline{Q}}^{\overline{#1}}}
\newcommand{\zjb}[1]{{\overline{J}}^{\overline{#1}}}
\newcommand{\zhb}[1]{{\overline{H}}^{\overline{#1}}}
\newcommand{\zshb}[1]{{\overline{h}}^{\overline{#1}}} 

\newcommand{\zrab}{R^{a\overline{a}}_{b\overline{b}}}
\newcommand{\ztaa}{T^{a\overline{a}}}

\newcommand{\zx}{\xi}
\newcommand{\ze}{\eta}
\newcommand{\zxb}{\overline{\xi}}
\newcommand{\zeb}{\overline{\eta}}

\newcommand{\zrb}{\rho}
\newcommand{\db}{\overline{d}}
\newcommand{\zw}{\omega}
\newcommand{\zt}{\theta}

\newcommand{\zsa}{|1\rangle}
\newcommand{\zsb}{|2\rangle}
\newcommand{\zsc}{|3\rangle}
\newcommand{\zsd}{|4\rangle}
\newcommand{\zsi}[1]{|{#1}\rangle}

\newcommand{\zD}{\Delta}

\newcommand{\ztp}[2]{{|{#1}\rangle\otimes |{#2}\rangle}}
\newcommand{\zsj}[1]{{\langle {#1} |}}

\newcommand{\cp}{{\cal P}}
\newcommand{\cs}{{\cal S}}
\newcommand{\wt}{\tilde{W}}
\newcommand{\vh}{\hat{v}}

\newcommand{\fb}{\frac{1}{\sqrt{2}}}
\newcommand{\fc}{\frac{1}{\sqrt{ q + q^{-1}}}}
\newcommand{\fd}{\frac{1}{\sqrt{2(q^2 + q^{-2})}}}

\newcommand{\lb}{\left(}
\newcommand{\rb}{\right)}
\newcommand{\zl}{\frac{1}{\gah}}
\newcommand{\ztf}{\frac{i\theta}{\pi}}
\newcommand{\zG}{\Gamma}
\newcommand{\ep}{\epsilon}


\title{The Exact \zss-Matrix for an \osp\ Disordered System}

\author{Zorawar S. Bassi\thanks{zorawar@mail.lns.cornell.edu} \\
Andr\'e LeClair \\ \\ Newman Laboratory \\ Cornell University \\
Ithaca, NY 14853, USA}

\date{December 1999}

\maketitle
\thispagestyle{empty}


\begin{abstract}
We study a two-dimensional disordered system consisting of Dirac fermions
coupled to a scalar potential.  This model is closely related to a more
general disordered system that has been introduced in conjunction with the
integer quantum Hall transition.  After disorder averaging, the interaction
can be written as a marginal \osp\ current-current perturbation.
The \osp\ current-current model in turn can be viewed as the fully 
renormalized version of an \ospa\ Toda-type system (at the marginal point).
We build non-local charges for the Toda system satisfying the 
\uospa\ quantum superalgebra.  The corresponding quantum group symmetry is
used to construct a Toda \zss-matrix for the vector representation.
We argue that in the marginal (or rational) limit, this \zss-matrix
gives the exact (Yangian symmetric) physical \zss-matrix for the fundamental
``solitons'' of the \osp\ current-current model. 
\newline
\vspace*{-0.1 in}
\flushleft{{\normalsize{CLNS 99/1646;\, hep-th/9911105}}}
\end{abstract}

\newpage

\zeqnorg%

\setcounter{section}{1}
\setcounter{equation}{0}

\section*{1. Introduction}

Non-perturbative techniques have proven to be very useful in the 
investigation of interacting field theories.  This is especially true in
two-dimensions, where rich symmetry structures have allowed the calculation
of fundamental quantities, such as exact \zss-matrices and correlation
functions, for many important models.  In recent times, this progress
has been further advanced by a better understanding of the mathematical
framework underlying the symmetry structures.  Perhaps this is most apparent
in the construction of exact \zss-matrices for integrable 2D systems.  Here
a knowledge of the symmetry algebra and its representation theory is
essential to building a \zss-matrix satisfying the Yang-Baxter constraint 
\cite{yangR,baxterR,jimboybR,jimboR,kulishR,sklyaninR}.
Well known examples of models with 
exact \zss-matrices include the sine-Gordon/massive Thirring
model \cite{zzsgR,abzsgR,thunR}, Gross-Neveu models 
\cite{karowskiR,andreiR,zzgnR}, principal chiral and sigma-type models 
\cite{wiegR,ogiR,reshiR,zzssmtR}, various statistical systems, 
such as RSOS systems \cite{andrewsR,ztbarsosR}, and Toda theories 
based on Lie algebras 
\cite{bradenR,christeR,grisaruR,holloheckeR,gandenR,nakatsuR,fringR}.  
A  relatively new class of models that
have been investigated are based on Lie superalgebras.  Supersymmetric Toda
models belong to this category and their exact \zss-matrices have been
calculated \cite{penatiR,evansR}.

In this paper we use the \zss-matrix approach to study a model based on the
Lie superalgebra \osp, which is closely related to a disordered system
introduced to describe the integer quantum Hall transition \cite{ludwigR}.  
As shown by Bernard \cite{berR}, the model arises after disorder averaging 
over a random scalar potential and
consists of a free fermionic and bosonic piece, combined with a marginal \osp\ 
current-current perturbation.  The bosonic part is the result of rewriting 
the fermion partition function as a path integral over complex bosonic
variables.  This pairs the fermions and bosons thus making the action 
``supersymmetric''.  The model is integrable with factorized scattering
and Yangian symmetry.  By introducing an anisotropy, which allows us to 
flow between a relevant and marginal perturbation, we construct non-local
charges for half the current-current operators.  The remaining current-current
operators are generated under renormalization at the marginal point.
The non-local charges are shown to satisfy the \uospa\ quantum superalgebra
and to be conserved to lowest order in conformal perturbation theory. 
Requiring the theory to have \uospa\ quantum
group symmetry, we calculate the \zss-matrix, or more appropriately the 
\zrr-matrix, for the fundamental vector representation.  The physical
\zss-matrix is then obtained, up to CDD factors, by imposing the unitarity
and crossing constraints.  We propose that in the marginal limit this
\zss-matrix is the exact \zss-matrix (in the fundamental representation) with
Yangian symmetry for the \osp\ current-current model.  The particle spectrum
of the model and the corresponding \zss-matrix are massive.
In particular, this means that all states are Anderson localized.


The quantum Hall disordered system discussed in \cite{ludwigR} contains three
types of randomness.  Though the generic case, with all types of randomness
present, is not believed to be integrable, it is certainly possible that on
some submainifold in the three coupling parameter space the model can be
exactly solved.  Various such subspaces have already been investigated 
\cite{ludwigR,berR,wenR}.  An interesting subset was recently 
studied in \cite{guruswamyR} (see also \cite{leclairqheR}), 
where supersymmetric disorder averaging led to a $gl(N|N)$ current-current type
model for which exact correlation functions were computed.  Our \zss-matrix
analysis considers the situation where there is only one specific type of
disorder, namely a random scalar potential.

We present our results as follows.  In section 2 we write down the models
and show how the full current algebra is generated under renormalization.
The non-local charges are constructed in section 3. {}From the quantum group
structure we build the \zss-matrix for the fundamental representation in
section 4.  The pole structure is briefly discussed in section 5.  Lastly we
conclude with a summary and comment on open questions for further study.
An appendix reviews the \osp\ algebras.

\addtocounter{section}{1}
\setcounter{equation}{0}

\section*{2. The \osp\ field theory models}

{\it 2a. The \osp\ current-current model}

\medskip
The model introduced by Ludwig {\it et al}.\ in connection with
the integer quantum Hall transition consists of a Dirac fermion 
$(\zpl,\zpr)$ coupled to, in the most general case,
three types of randomness: a random vector potential $A$, a random mass $m$,
and a random scalar potential $V$.  The Euclidean action takes the 
form $(z=t+ix)$
\zbe
 S  =  \int {d^2 x \over 2\pi} \, \Bigl( \psm (\zd_{\zb} - i A_{\zb})\pp +
\pbm(\zd_z - i A_z)\pbp  
+ i \frac{m(x)}{2}(\pbm \pp - \psm \pbp) 
+ i \frac{V(x)}{2}(\pbm \pp + \psm \pbp) \Bigr). \label{disacgenZ}
\zee
The model we are interested in only contains a random scalar potential
$(A=m=0)$, which is taken to have a gaussian distribution with mean zero and 
positive variance $g_V$
\zbe
P[V] = \exp\left( - \frac{1}{4 g_V}\int \zm\, V(x)^2 \right).\label{gdistVZ}
\zee
Using the supersymmetric method and averaging over $V(x)$ leads to the
effective action \cite{berR}
\zbe
S_{\rm eff} = S_{\rm cft} + \frac{g_V}{4} \int \zm\, \Phi_V \label{seffVZ}
\zee
\zbe
S_{\rm cft} =  \int \zm \, \left( \psm\zd_\zb \pp + \pbm\zd_z \pbp +
\bt\zd_\zb \ga + \btb\zd_z\gab \right) \label{scftZ}
\zee
\zbe
\Phi_V = \left( \pbm\pp + \psm\pbp + \btb\ga + \gab\bt\right)^2, \label{intVZ}
\zee
where $(\bt,\ga,\btb,\gab)$ are complex bosonic ghosts of conformal
dimensions
\zbe
[\bt] = (\fa,0),\ \  [\ga] = (\fa,0), \ \ [\btb] = (0,\fa), \ \
[\gab] = (0,\fa), \label{bdimZ}
\zee
and satisfy the operator product expansions (OPE's)
\zbe
\ga(z)\bt(w) \sim - \bt(z)\ga(w) \sim \frac{1}{z-w}. \label{bopeZ}
\zee
The fermions satisfy the usual OPE's
\zbe
\pp(z)\psm(w) \sim \psm(z)\pp(w) \sim \frac{1}{z-w}. \label{fopeZ}
\zee
The action $S_{\rm cft}$ is anti-hermitian and has a central charge 
of zero, $c=0$, since $c_{\bt,\ga} = -1$.  
One can view $S_{\rm eff}$ as a perturbed (non-unitary) conformal 
field theory (CFT). Note that the perturbation is marginal ($\Phi_V$
has dimension 2). 
We will see below that the perturbation is actually marginally relevant.

As in the random bond Ising model 
\cite{berR,shankarR,dotsenkoR,simonettiR,cabraR}, 
the following supersymmetric current
algebra is a symmetry of the conformal action $S_{\rm cft}$ 
\[
 G_\pm(z) = \bt(z) \zpl(z), \quad  \gh_\pm(z) = \ga(z)\zpl(z)
\]
\[
 K(z) = :\bt^2(z):, \quad  \kh(z) = :\ga^2(z):
\]
\zbe
 J(z) = :\pp(z)\psm(z):, \quad H(z) = :\ga(z)\bt(z): \label{holcuZ}
\zee
\[
 \gb_\pm(\zb) = \pm\btb(\zb) \zpr(\zb), \quad  
 \ghb_\pm(\zb) = \mp\gab(\zb)\zpr(\zb)
\]
\[
 \kb(\zb) = -:\btb^2(\zb):, \quad  \khb(\zb) = -:\gab^2(\zb):
\]
\zbe
 \jb(\zb) = :\pbp(\zb)\pbm(\zb):, \quad 
 \hb(\zb) = :\gab(\zb)\btb(\zb):. \label{aholcuZ}
\zee
(In this paper we use the term supersymmetry merely to mean a symmetry
algebra/transformation based on some Lie superalgebra and not space-time
supersymmetry in the usual sense.)  
Here $:\ldots :$ denotes normal ordering, namely $:A(w) B(w):$ is the
coefficient of $(z-w)^0$ in the OPE $A(z)B(w)$.
These currents form a level one
representation of the affine \ospa\ current algebra.  The rank two \osp\ Lie
superalgebra has six roots, four odd or ``fermionic'' and two
even or ``bosonic''.
The $G$'s, being fermionic, are associated with the fermionic roots and the
$K$'s with the bosonic roots.  The two simple roots, $(\al_1,\al_2)$ 
can be chosen to be both fermionic or one fermionic and one bosonic.  
We will choose a purely fermionic simple root system.  With this choice the 
additional affine root, $\al_0 = -(\al_1+\al_2)$, is associated with $K(z)$ 
(or $\kh(z)$).  (The \osp\ algebras are discussed in the appendix.)
The non-trivial OPE's for the currents are
\[
 J(z)J(w) \sim \frac{1}{(z-w)^2}, \quad H(z)H(w) \sim \frac{-1}{(z-w)^2}
\]
\[
J(z)G_\pm(w) \sim \frac{\pm 1}{(z-w)}G_\pm(w), \quad
J(z)\gh_\pm(w) \sim \frac{\pm 1}{(z-w)}\gh_\pm(w)
\]
\[
H(z)G_\pm(w) \sim \frac{1}{(z-w)}G_\pm(w), \quad
H(z)\gh_\pm(w) \sim \frac{-1}{(z-w)}\gh_\pm(w)
\]
\[
H(z)K(w) \sim \frac{2}{(z-w)}, \quad H(z)\kh(w) \sim \frac{-2}{(z-w)}
\]
\[
 \gh_\pm(z) G_\mp(w) \sim \frac{1}{(z-w)^2} + \frac{1}{(z-w)} (H(w)\pm J(w))
\]
\[
\kh(z) K(w) \sim \frac{2}{(z-w)^2} + \frac{4}{(z-w)}H(w)
\]
\[
G_-(z)G_+(w) \sim \frac{1}{(z-w)} K(w), \quad 
\gh_-(z) \gh_+(w) \sim \frac{1}{(z-w)}\kh(w)
\]
\zbe
K(z)\gh_\pm(w) \sim \frac{-2}{(z-w)} G_\pm(w), \quad
\kh(z)G_\pm(w) \sim \frac{2}{(z-w)}\gh_\pm(w), \label{holopeZ}
\zee
and similarly for the anti-holomorphic currents.
We want to emphasize
that even though some of the anti-holomorphic currents (\ref{aholcuZ})
differ from their holomorphic counterparts (\ref{holcuZ}) by signs 
(e.g.\ $K = :\bt^2:$ but $\kb = - :\btb^2:$), their OPE's are nevertheless the
same, i.e., the usual OPE's obtained from (\ref{holopeZ}) by replacing all
operators ${\cal O}(z)$ by $\overline{{\cal O}}(\zb)$.

The field $\Phi_V$ can now be written as
\zbe
\Phi_V = -2 \left[ \jb J - \hb H + \fa(\kb \kh + \khb K) 
+ \gb_- \gh_+ - \ghb_- G_+ + \gb_+ \gh_- - \ghb_+ G_- \right],
\label{intVcuZ}
\zee
which is of the current-current form (see appendix).  The interaction is thus
a current-current perturbation that preserves the \osp\ symmetry
of $S_{\rm cft}$.  Furthermore, this implies that $S_{\rm eff}$ has
Yangian symmetry \cite{berR}.  Any \zss-matrix we construct must respect this
Yangian symmetry.  

The operator $\Phi_V$ alone forms a closed algebra.  Its OPE can be
calculated using (\ref{bopeZ}) and (\ref{fopeZ}) (or (\ref{holopeZ}) and
its anti-holomorphic version), and is found to be
\zbe
 \Phi_V(z,\zb) \Phi_V(w,\wb) \sim - \frac{8}{|z - w|^2} \Phi_V(w,\wb).
 \label{intVopeZ}
\zee
This leads to the beta function (to lowest order)
\zbe
\bt_{g} = \frac{dg}{d\log R} = g^2, \label{betaVZ}
\zee
where $R$ is a length scale (see section 2b) and henceforth we drop the
$V$ subscript on $g_V$, writing $g\equiv g_V$.  We see that the perturbation
is marginally relevant: $g$ (which is positive by definition)
increases at large distances.  The theory is asymptotically free in the UV.
The model $S_{\rm eff}$ is thus in a massive regime and the \zss-matrix
we calculate will describe scattering of massive particles.  The behaviour
here is opposite to that of the random bond Ising/random mass model.
The random mass model ($A=V=0$) perturbing field $\Phi_M$, with
coupling $g_M$, can also be 
written as a current-current perturbation \cite{berR}.  This 
requires redefining four of the \osp\ currents as
\zbe
\kb \rightarrow -\kb,\ \ \khb \rightarrow -\khb, \ \ 
\gb_- \rightarrow -\gb_-, \ \ \ghb_+ \rightarrow -\ghb_+, \label{redefcuZ}
\zee
with all other currents unchanged.  The operator algebra is unchanged under
this transformation.  One finds that
\zbe
\Phi_M(z,\zb) = - \Phi_V(z,\zb). \label{intMZ}
\zee
Due to the minus sign, the beta function changes sign
\zbe
\bt_{g_M} = - g_M^2, \label{betaMG}
\zee
giving a marginally irrelevant perturbation that is asymptotically free in
the infra-red.  In this case one expects to flow to a massless regime.
(For a non-perturbative analysis of the random bond Ising model, 
including a discussion of the \zss-matrix, see \cite{simonettiR,cabraR}.) 

\medskip
\noindent
{\it 2b. The \ospa\ Toda-type model}

\smallskip
One can try to construct the Yangian charges for (\ref{seffVZ}) and then use
these to build the \zss-matrix.  However we will follow an alternative approach
along the lines of \cite{leclairR}, which though being 
less direct, is easier to apply.

In \cite{leclairR} it is shown that the Yangian symmetry associated with a
current-current perturbation of the WZW model 
(based on a simply laced Lie algebra) can
be extracted (as a marginal limit) from a ``smaller'' model.
This smaller model, which is simply a Toda
system with imaginary coupling, contains only current-current operators
corresponding to the affine simple roots.  The reason for working with
this smaller model is that one can construct an affine quantum 
group ($q$-deformed) 
structure, using the method of non-local charges, only for the Toda system.  
The full current-current perturbation
is recovered through renormalization of the Toda model (see below).
{}From the affine quantum group symmetry 
the \zss-matrix can be constructed.  The
marginal limit of the Toda \zss-matrix then leads to the Yangian symmetric
\zss-matrix for the original current-current model.  

We now apply the same reasoning to the \osp\ model.  As in the Lie
algebra case, quantum group charges cannot be constructed for the full 
model (\ref{seffVZ}).  A smaller model which does have quantum group
symmetry, as we will show in the next section, consists of
taking only half the terms in (\ref{intVcuZ}).  Its action takes the form
\zbe
\tilde{S} = S_{\rm cft} + \frac{g}{4}  \int\zm\, \tilde{\Phi}_V,
\label{todaacZ}
\zee
with $S_{\rm cft}$ as in (\ref{scftZ}) and
\zbe
 \tilde{\Phi}_V = -2 \left[ \fa \khb K + \gb_- \gh_+ + \gb_+ \gh_-\right].
\label{intTZ}
\zee
The current terms retained correspond to the affine simple roots of \ospa .
We have also dropped the Cartan terms $\jb J$ and $\hb H$.
Thus (\ref{todaacZ})
is the supersymmetric analog of the Toda system 
used in \cite{leclairR} to study
current-current perturbations of the WZW model.  (In order to construct the
quantum charges one needs to introduce an additional parameter $\bh$ in
(\ref{intTZ}), which serves to make the perturbation relevant.  
The model (\ref{todaacZ}) is then understood 
as the marginal limit $\bh \rightarrow 1$ of the deformed model.  This will
become clearer in the next section.)

Unlike the full model, $\tilde{S}$ is not renormalizable.  The reason
being that the operators in (\ref{intTZ})
do not form a closed algebra by themselves.
More generally, consider a marginal perturbation of a CFT
by some set of operators $\{\zo^i(z,\zb)\}$
\zbe
S = \scft + \sum_i \int d^2x\, g_i \zo^i(z,\zb). \label{gpcftZ}
\zee
For (\ref{gpcftZ}) to be renormalizable, the operator algebra for the set
$\{\zo^i\}$ must close onto itself.  That is to say the OPE's must be
of the form
\zbe
\zo^i(z,\zb)\zo^j(w,\wb) \sim C^{ij}_k  \frac{1}{|z-w|^2} \zo^k(w,\wb),
\label{genopeZ}
\zee
for some structure constants $C^{ij}_k$, with any operators appearing in
(\ref{genopeZ}) already being present in the action (\ref{gpcftZ}).  In this
case Zamolodchikov \cite{zamobfR} has shown that the 
beta functions to lowest order
(1 loop) are
\zbe
\bt_{g_i} = \frac{d g_i}{d\log R} = -\pi C_i^{jk} g_j g_k. \label{genbetaZ}
\zee
Specializing to (\ref{todaacZ}), the action can be written as
\zbe
\tilde{S} = \scft + \sum_{i=1}^3 g_i \int d^2x\, \zo^i, \label{genTacZ}
\zee
where
\zbe
\zo^1 = \gb_-\gh_+, \quad \zo^2 = \gb_+ \gh_-,\quad \zo^3=\khb K,
\label{op1Z}
\zee
and we have allowed the couplings to be independent.  Running the
renormalization procedure, one finds that after the first iteration the
following operators are generated
\zbe
\zo^4 = \ghb_+G_-, \quad \zo^5 = \ghb_- G_+,\quad \zo^6=\kb\kh.
\label{op2Z}
\zee
These correspond to the negative simple roots.  However, the operator algebra
for $\{\zo^i\}_{1\leq i\leq 6}$ still does not close.  A second run generates
the Cartan operators
\zbe
\zo^7 = (\hb +\jb)(H+J), \quad \zo^8 =(\hb -\jb)(H-J),\quad\zo^9=\hb H.
\label{op3Z}
\zee
The operator algebra for $\{\zo^i\}_{1\leq i\leq 9}$ closes, and the 
resulting action is
\zbe
\tilde{S} = \scft +\int d^2x\, \sum_{i=1}^9 g_i \zo^i.
\zee
The beta functions are found using (\ref{genbetaZ}) to be
\[ \bt_{g_1} = 2 \pi g_1 (4 g_8 + g_9) + 8 \pi g_5 g_6,\qquad 
   \bt_{g_2} = 2 \pi g_2 (4 g_7 + g_9) + 8 \pi g_4 g_6 \]
\[ \bt_{g_3} = 8 \pi g_3 (g_7 + g_8 + g_9) + 2 \pi g_4 g_5 \]
\[ \bt_{g_4} = 2 \pi g_4 (4 g_8 + g_9) + 8 \pi g_2 g_3, \qquad
   \bt_{g_5} = 2 \pi g_5 (4 g_7 + g_9) + 8 \pi g_1 g_3 \]
\[ \bt_{g_6} = 8 \pi g_6 (g_7 + g_8 + g_9) + 2 \pi g_1 g_2 \]
\[ \bt_{g_7} = 2 \pi g_1 g_4, \qquad
   \bt_{g_8} = 2 \pi g_2 g_5 \]
\zbe 
\bt_{g_9} = 32 \pi g_3 g_6. \label{beta9Z}
\zee
We can easily verify that a solution to (\ref{beta9Z}) exists which gives
the original model (\ref{seffVZ}) and its beta function (\ref{betaVZ}).
Since (\ref{seffVZ}) contains only one coupling, we should try the ansatz
$ g_i  = -\frac{1}{4\pi} \al_i g $,
where $\al_i$ are some numerical constants.  (We remark that the $\al_i$'s do
not spoil the quantum group symmetry and one can work with
(\ref{genTacZ}) using this ansatz
instead of (\ref{todaacZ}).)  It is straight-forward to check that the set
\zbe
\{\al_1=\al_2=1,\al_3 =\fa,\al_4=\al_5=-1,\al_6=\fa,\al_7=\al_8=\fa,
\al_9=-2\}, \label{alphasetZ}
\zee
reproduces (\ref{seffVZ}) and reduces (\ref{beta9Z})
to the single beta function (\ref{betaVZ}) for $g$.
Thus under renormalization (\ref{todaacZ})
leads to the \osp\ current-current model.  We
will therefore construct the \zss-matrix for the Toda-type system
(\ref{todaacZ}) and argue that it gives the
required \zss-matrix for (\ref{seffVZ}) in the marginal limit.

\addtocounter{section}{1}
\setcounter{equation}{0}

\section*{3. Quantum group symmetry in the \osp\ models}

{\it 3a. Non-local charges and conformal perturbation theory}

\smallskip
Quantum group symmetry is realized by non-local charges 
\cite{leclairR,berhyR,luscherR,vegaR} constructed using
conformal perturbation theory.  We outline the main points of the
construction below.  (For more details see \cite{leclairR}.)

Suppose we have a CFT perturbed by a relevant spin-zero
field
\zbe \zpp(z,\zb) = \zps(z)\zpsb(\zb), \zee
with the Euclidean action
\zbe 
S = \scft + g \int \zm\, \zpp(z,\zb), \label{gpertacZ} 
\zee
and some currents $\{J^a,\zjb{a}\}$ which are chiral when $g=0$
\zbe
\zdz J^a = \zdb \zjb{a} = 0 \Longrightarrow J^a=J^a(z), \ \ 
\zjb{a}=\zjb{a}(\zb).
\zee
In the perturbed theory (\ref{gpertacZ}) these currents are no longer chiral,
but to lowest order in $g$ satisfy the following 
equations of motion \cite{zamopcftR}
\[
\zdb J^a(z,\zb) = g \oint_z \zmw \zpp(w,\zb) J^a(z)
\]
\zbe
\zdz \zjb{a}(z,\zb) = g \oint_\zb \zmwb \zpp(z,\wb) \zjb{a}(\zb). 
\label{cueomZ}
\zee
If the residues of the OPE's on the righthand side are total derivatives
\zbe
{\rm Res}_{z=w}(\zps(w)J^a(z)) = \zdz h^a(z), \quad
{\rm Res}_{\zb=\wb}(\zpsb(\wb)\zjb{a}(\zb)) = \zdb\zshb{a}(\zb),
\label{gresZ}
\zee
then (\ref{cueomZ}) becomes
\zbe
\zdb J^a(z,\zb) = \zdz H^a(z,\zb),\quad
\zdz \zjb{a}(z,\zb) = \zdz \zhb{a}(z,\zb), \label{cuceomZ}
\zee
where
\zbe
H^a(z,\zb) = g h^a(z)\zpsb(\zb),\quad
\zhb{a}(z,\zb) = g \zshb{a}(\zb)\zps(z). 
\zee
The equations of motion (\ref{cuceomZ}) imply that to lowest order
in $g$ we have the conserved charges
\[
Q^a = \frac{1}{2\pi i}\left( \int dz\, J^a + \int d\zb\, H^a\right)
\]
\zbe
\zqb{a} = \frac{1}{2\pi i}\left( \int d\zb\, \zjb{a} + \int dz\, \zhb{a} 
\right). \label{ccrgZ}
\zee

The currents $\{ J^a,H^a\}$ and $\{\zjb{a},\zhb{a}\}$, and hence
the charges $\{Q^a,\zqb{a}\}$, are non-local.  To see this consider the
specific case where $\scft$ is a sum of free bosons
\zbe
\scft = \frac{1}{8\pi} \int d^2x\, \sum_i \zd_\mu \Phi_i \zd_\mu \Phi_i.
\label{bcftZ}
\zee
In the limit $g=0$, $\Phi_i$ can be expanded into its chiral components,
$\Phi_i(z,\zb)=\zp_i(z) +\zpb_i(\zb)$, with
\zbe 
\langle\zp_i(z)\zp_j(w)\rangle = -\delta_{ij}\log(z-w),\quad
\langle\zpb_i(\zb)\zpb_j(\wb)\rangle = -\delta_{ij}\log(\zb-\wb).
\label{bcorZ}
\zee
If $g\neq 0$, $\Phi_i$ can again be written as 
$\Phi_i(x,t) = \zp_i(x,t)+\zpb_i(x,t)$, but now $\zp_i(x,t)$ and 
$\zpb_i(x,t)$ are no longer chiral.  For arbitrary $g$, $\zp_i(x,t)$ and
$\zpb_i(x,t)$ can be written in the non-local way
\[
\zp_i(x,t) = \fa\left( \Phi_i(x,t) + \int_{-\infty}^x dy\zd_t\Phi_i(y,t)
\right) \stackrel{g=0}{=} \zp_i(z)
\]
\zbe
\zpb_i(x,t) = \fa\left( \Phi_i(x,t) - \int_{-\infty}^x dy\zd_t\Phi_i(y,t)
\right) \stackrel{g=0}{=} \zpb_i(\zb). \label{nlchbZ}
\zee
Since the currents are in general functions of $\zp_i$ and $\zpb_i$, they 
are non-local due to (\ref{nlchbZ}).  This non-locality leads to non-trivial
braiding relations for the currents
\zbe
J^a(x,t) \zjb{a}(y,t) = \zrab \zjb{b}(y,t) J^b(x,t), 
\zee
where $\zrab$ is a braiding matrix which will depend on the couplings and the
parities of the currents, i.e.\ whether the currents are even (bosonic)
or odd (fermionic).  The corresponding result for the charges is
\zbe
Q^a \zqb{a} - \zrab \zqb{b} Q^b = \ztaa,  \label{gcrgbrrZ}
\zee
where $\ztaa$ is a topological charge
\zbe
\ztaa = \frac{g}{2\pi i} \int_t (dz\zdz + d\zb\zdb) h^a(z)\zshb{a}(\zb).
\label{gtcrgZ}
\zee

\medskip
\noindent
{\it 3b. Bosonization and the $\bh$ parameters}

\smallskip
The above formalism is valid for a relevant perturbation.  However, the
perturbing field (\ref{intTZ}) is marginal.  By making the currents depend
on a parameter $\bh$, the perturbation can be made
relevant, with the marginal limit being $\bh\rightarrow 1$.  In
order to introduce this additional parameter the action (\ref{todaacZ})
has to be (partially) bosonized.

The fermions can be bosonized in the standard way
\zbe
\zpl(z) = e^{\pm i\zp_1(z)},\quad \zpr(\zb) = e^{\mp i\zpb_1(\zb)}.
\zee
The bosonic ghosts can be written as \cite{friedanR,ketovR}
\[
\ga(z) = e^{\zp_2(z)} \ze(z),\quad 
\gab(\zb) = e^{-\zpb_2(\zb)} \zeb(\zb)
\]
\zbe
\bt(z) = e^{-\zp_2(z)} \zd\zx(z), \quad
\btb(\zb) = e^{\zpb_2(\zb)} \zdd\zxb(\zb).
\zee
The chiral bosons $\{ \zp_i(z),\zpb_i(\zb)\}$ satisfy (\ref{bcorZ}), and 
$\{ \ze(z),\zx(z)\}$ and $\{\zeb(\zb),\zxb(\zb)\}$ are fermionic ghost
systems with conformal dimensions and OPE's
\zbe
[\ze]=(1,0),\ \ [\zx]=(0,0),\ \ [\zeb]=(0,1),\ \ [\zxb]=(0,0)
\zee
\zbe
\ze(z)\zx(w)\sim\zx(z)\ze(w)\sim\frac{1}{(z-w)},\quad
\zeb(\zb)\zxb(\wb)\sim\zxb(\zb)\zeb(\wb)\sim\frac{1}{(\zb-\wb)}.
\zee
The central charge for the ghost system, $c_{\ze,\zx}$, is $-2$.
The \ospa\ currents expressed in terms of the new fields become 
\[
J = i\zd\zp_1,\quad \jb = -i\zdd\zpb_1
\]
\[
H = \zd\zp_2,\quad \hb = -\zdd\zpb_2
\]
\[
G_\pm = \exp\left(i\al_{1,2}\cdot\zpv\right) \zd\zx, \quad
\gb_\pm = \pm \exp\left(-i\al_{1,2}\cdot\zpbv\right) \zdd\zxb
\]
\[
\gh_\pm = \exp\left(-i\al_{2,1}\cdot\zpv\right) \ze, \quad
\ghb_\pm = \mp \exp\left(i\al_{2,1}\cdot\zpbv\right)  \zeb
\]
\[
K = \exp\left(-i\al_0\cdot\zpv\right) :\zd^2\zx\,\zd\zx:,\quad
\kb = -\exp\left(i\al_0\cdot\zpbv\right) :\zdd^2\zxb\,\zdd\zxb:
\]
\zbe
\kh = \exp\left(i\al_0\cdot\zpv\right) :\zd\ze\,\ze:,\quad
\khb = -\exp\left(-i\al_0\cdot\zpbv\right):\zdd\zeb\,\zeb:,
\label{boscurelZ}
\zee
where 
\zbe
\zpv = (\zp_1,\zp_2),\ \  \zpbv = (\zpb_1,\zpb_2),
\ \ \zPv=(\Phi_1,\Phi_2),
\zee
and we have introduced the simple roots for \ospa\ (see appendix)
\zbe
\al_1=(1,i),\ \ \al_2 = (-1,i), \ \  \al_0=(0,-2i). \label{srootsZ}
\zee
In deriving the expressions for $H, K$ and $\kh$, we have made use of
\zbe
e^{\pm\zp_2(z)} e^{\mp\zp_2(w)} = (z-w)(1\pm\zd_w\zp_2(w)(z-w)+\ldots)
\zee
\zbe
e^{\pm\zp_2(z)} e^{\pm\zp_2(w)} = \frac{1}{(z-w)}e^{\pm 2\zp_2(w)} +\ldots
\zee
\zbe
\ze(z)\ze(w) = (z-w)(:\zd_w\ze(w)\ze(w): + \ldots)
\zee
\zbe
\zdz\zx(z)\zd_w\zx(w) = (z-w)(:\zd_w^2\zx(w)\zd_w\zx(w): +\ldots )
\zee
\zbe
:\ze(w)\ze(w):\,  =\,  :\zd_w\zx(w)\zd_w\zx(w):\,  = 0,
\zee
and similarly for the anti-holomorphic parts.  One can check that
(\ref{boscurelZ}) satisfy the required OPE's (\ref{holopeZ}).

Now we can write the perturbed CFT
which renormalizes to (\ref{seffVZ}) at the marginal point.  
The required action takes the form
\zbe
\tilde{S}^{d} = \scft + S_{\rm pert.}^d \label{dtacZ}
\zee
\zbe
\scft = \frac{1}{4\pi} \int d^2x\,\left( \fa(\zd_\mu \Phi_1)^2 +
\fa(\zd_\mu \Phi_2)^2 + \ze\zdd\zx + \zxb\zd\zeb \right)
\label{bscftZ}
\zee
\zbe
S_{\rm pert.}^d = -\frac{g}{4} \int d^2x \left(
e^{-i\bh \al_1\cdot\zPv} \zdd\zxb\ze - 
e^{-i\bh \al_2\cdot\zPv} \zdd\zxb\ze -
\fa e^{-i\bh \al_0\cdot\zPv}:\zdd\zeb\,\zeb:\,:\zd^2\zx\,\zd\zx : 
\right). \label{dintVZ}
\zee
At the marginal point, $\bh=1$, $\tilde{S}^d$ gives the 
bosonized version of $\tilde{S}$.  
The beta parameter is analogous to the $\bt$ (or $\bh$) parameter 
in the sine-Gordon model \cite{zzsgR,leclairR}, and 
can be thought of as arising due to an anisotropy in
the current-current perturbation.  If we add to (\ref{todaacZ}) the term
\zbe 
\frac{1}{8\pi}\int d^2x\,\rho \left(  \jb J -  \hb H \right) =
\frac{1}{8\pi}\int d^2x \,\left( \rho(\zd_\mu\Phi_1)^2 + 
\rho(\zd_\mu\Phi_2)^2 \right),
\zee
and rescale $\Phi_i \rightarrow \Phi_i/\sqrt{1+\rho}$, one arrives at
(\ref{dtacZ}) with $\bh = 1/\sqrt{1+\rho}$.  In
view of this one can interpret $\ztS^d$ as a deformation of (\ref{todaacZ}).
Indeed we will show that the beta parameter
is related to the quantum group deformation parameter.
The Toda-type structure is clearly seen in the bosonized form (\ref{dintVZ}).
In comparison with the Lie algebraic Toda
models, the main difference here is the inclusion of fermionic fields
due to supersymmetry.  

\ignorethis{For arbitrary positive values of the beta parameters, the action $\ztS^d$ 
is inconsistent.  The first two terms in (\ref{dintVZ}) imply
that the coupling $g$ has the dimensions 
\zbe 
\Delta(g) = \overline{\Delta}(g) = \frac{\bh_2^2}{2} - \frac{\bh_1^2}{2},
\zee
whereas the remaining $\khb K$ term gives
\zbe
\Delta(g) = \overline{\Delta}(g) = 2\bh_2^2 - 2.
\zee
Consistency thus requires $\bh_1$ and $\bh_2$ satisfy the constraint
\zbe
  3 \bh_2^2 + \bh_1^2 = 4. \label{gconsZ}
\zee
Of course the marginal point lies on this curve.  The perturbation
(\ref{dintVZ}) is relevant for the regime
\zbe
   \bh_2 > \bh_1, \quad \bh_2 > 1,  \label{gregZ}
\zee
and we can use the previously discussed method of non-local charges to
construct the quantum charges.  Then in the marginal limit 
$\bh_i\rightarrow 1$ subject to the constraint (\ref{gconsZ}), the Yangian 
structure of (\ref{todaacZ}), and hence that of (\ref{seffVZ}),
can be obtained.}

For arbitrary positive values of $\bh$, the three terms in $S_{\rm pert.}^d$ do not
have the same dimensions.  The first two terms are marginal for all values of
$\bh$, whereas the remaining $\khb K$ term is relevant for $\bh> 1$.  This 
suggests that $S_{\rm pert.}^d$ should be rewritten with two couplings, 
$g_1$ and $g_2$, with one of them being dimensionless, which in turn seems to
imply that conformal perturbation theory 
cannot be used to construct the non-local charges.  However, these problems 
can be resolved by introducing a background charge $\zq$ coupled to 
the field $\Phi_2$ \cite{ketovR}. The conformal dimensions of 
$\exp(i\al\zp_2)$ and $\exp(i\al\zpb_2)$ are then changed from $\al^2/2$ to
$\al(\al-\zq)/2$.  One finds that for a purely imaginary charge of
\zbe
\zq = i \frac{4}{3} \frac{\bh^2 -1}{\bh}, \label{backcrgZ}
\zee
all three terms take of the same conformal dimensions and giving
\zbe
\zD(g) = \overline{\zD}(g) = \frac{2}{3}(\bh^2 - 1). \label{cfdimgZ}
\zee
Therefore the perturbation as it is written in (\ref{dintVZ}), with one coupling $g$, 
is consistent and can be treated as being relevant for $\bh>1$.  Since introducing
$\zq$ does not change the OPE's, the construction of the non-local charges is identical
for both cases $\zq=0$ and $\zq\neq 0$.  We choose to work with $\zq=0$.  All 
expressions that follow, in particular the quantum group symmetry equations, also
hold for $\zq\neq 0$.  The only difference between $\zq=0$ and $\zq\neq 0$ is that
the spins of the charges are changed.  This means that in going from $\zq=0$ to 
$\zq\neq 0$, the gradation of the quantum affine algebra changes from the homogeneous
to the principal gradation (see below).  
Thus by taking $\zq=0$ we are effectively working in the homogeneous gradation.

\medskip
\noindent
{\it 3c. The quantum group charges}

\smallskip
In this section we construct the currents $\{J^a,H^a\}$ and 
$\{\zjb{a},\zhb{a}\}$, and the charges $\{Q^a,\zqb{a}\}$,
satisfying the conservation laws (\ref{cuceomZ}) for 
the theory (\ref{dtacZ}).

Taking into account (i) the Toda-type structure of (\ref{dintVZ}) and (ii)
the forms of the known currents for 
non-supersymmetric Toda \cite{leclairR}, one
expects the $J^a$'s to be some combinations of the vertex operators
with the fermions
\zbe
J^a \sim \exp\left(i a_1 \zp_1 + i a_2 \zp_2\right) \times
\{\ze,\zd\zx,\zd\ze\ze,\zd^2\zx\zx\},
\zee
and similarly for the $\zjb{a}$'s.  
\ignorethis{We consider the following set
\[
\tilde{J}^{1,2} = \exp\left( i \al_{1,2}\cdot\zpv_{d^\prime}\right) \zd\zx, 
\quad
\tilde{\jb}^{{}_{\scriptstyle 1,2}} = \exp\left( 
i \al_{1,2}\cdot\zpbv_{d^\prime}\right) \zeb
\]
\zbe
\tilde{J}^0 = \exp\left( i \al_0\cdot\zpv_{d^\prime}\right) :\zd\ze\,\ze:, 
\quad
\tilde{\jb}^{{}_{\scriptstyle 0}} = 
\exp\left( i \al_0\cdot\zpbv_{d^\prime}\right) 
:\zdd^2\zxb\,\zd\zxb:,
\zee
where
\zbe
\zpv_{d^\prime} = \left(\frac{\zp_1}{\bh_1},\frac{\zp_2}{\bh_2}\right), \ \
\zpbv_{d^\prime} = \left(\frac{\zpb_1}{\bh_1},\frac{\zpb_2}{\bh_2}\right),\ \
\zPv_{d^\prime} = \left(\frac{\Phi_1}{\bh_1},\frac{\Phi_2}{\bh_2}\right).
\zee
The expressions for $\tilde{J}^{1,2}$ and 
$\tilde{\jb}^{{}_{\scriptstyle 1,2}}$ are not
quite correct.
For each $\tilde{J}^i$ and $\tilde{\jb}^{{}_{\scriptstyle i}}$, 
the only non-vanishing contribution to (\ref{cueomZ}) comes
from the OPE with the $\al_i$ term in (\ref{dintVZ}).
Take for example $\tilde{J}^1$, with the other cases
being similar.  The OPE that needs to be evaluated is:
$\exp\left( -i\al_1\cdot\zpv_d(w)\right)\ze(w) \tilde{J}^1(z)$.  Using
Wick's theorem, and the standard OPE's for the vertex operators and 
fermions, we get
\begin{eqnarray}
\exp\left( -i\al_1\cdot\zpv_d(w)\right)\ze(w) \tilde{J}^1(z) & \sim  &
\frac{1}{(w-z)} \exp\left[ -i\left( \bh_1 - \frac{1}{\bh_1}\right) i\zp_1(z)
-i\left( \bh_2 - \frac{1}{\bh_2} \right) i\zp_2(z) \right] \nonumber \\
& & \mbox{} \times
\left( -i\bh_1 \zd_z\zp_1(z) - i \bh_2 i \zd_z\zp_2(z)\right).
\end{eqnarray}
This has the correct form (\ref{gresZ}), i.e.\ can be turned into a total
$\zd_z$ derivative, only if $\bh_1=\bh_2$.  Setting $\bh_1=\bh_2$ and
defining
\zbe 
\bh \equiv \bh_1 = \bh_2,
\zee
the right side reduces to
\zbe
\frac{1}{(w-z)} \zd_z \left( \gah e^{-i\frac{\bh}{\gah}\al_1\cdot\zpv(z)}
\right), \label{totderZ}
\zee
where
\zbe
\frac{1}{\gah} \equiv 1 - \frac{1}{\bh^2} \geq 0.  \label{defgaZ}
\zee
{}From (\ref{totderZ}) we get a conservation law of the form (\ref{cuceomZ}).
The same situation occurs for
$\tilde{J}^{2}$ and $\tilde{\jb}^{{}_{\scriptstyle 1,2}}$.  
In contrast, since the 
currents $\tilde{J}^0$ and $\tilde{\jb}^{{}_{\scriptstyle 0}}$ involve
only the parameter $\bh_2$, they are conserved independent of the
value of $\bh_1$.}
We find the following currents
\[
{J}^{1,2} = \exp\left( \frac{i}{\bh} \al_{1,2}\cdot\zpv\right) \zd\zx
\]
\[
{J}^0 = \exp\left( \frac{i}{\bh} \al_0\cdot\zpv\right) :\zd\ze\,\ze:
\]
\[
{\jb}^{1,2} = \exp\left(
\frac{i}{\bh} \al_{1,2}\cdot\zpbv \right) \zeb
\]
\zbe
{\jb}^{0} =
\exp\left( \frac{i}{\bh} \al_0\cdot\zpbv\right) 
:\zdd^2\zxb\,\zd\zxb:. \label{hahJZ}
\zee
\[
H^{1,2} =  \rho_{1,2}^{(0)}\,  g \gah \exp \left(
-i\frac{\bh}{\gah} \al_{1,2}\cdot\zpv - i\bh \al_{1,2}\cdot \zpbv\right)
\zdd\zxb
\]
\[
H^{0} =  \rho_{0}^{(0)}\, g \gah \exp \left(
-i\frac{\bh}{\gah} \al_{0}\cdot\zpv - i\bh \al_{0}\cdot \zpbv\right)
:\zdd\zeb\,\zeb:
\]
\[
\hb^{1,2} =  \zrb_{1,2}^{(0)}\, g \gah \exp \left(
-i\frac{\bh}{\gah} \al_{1,2}\cdot\zpbv - i\bh \al_{1,2}\cdot \zpv\right) \ze
\]
\zbe
\hb^{0} =  \zrb_{0}^{(0)}\, g \gah \exp \left(
-i\frac{\bh}{\gah} \al_{0}\cdot\zpbv - i\bh \al_{0}\cdot \zpv\right)
:\zd^2\zx\,\zd\zx:, \label{hahHZ}
\zee
where
\zbe
\frac{1}{\gah} \equiv 1 - \frac{1}{\bh^2} \geq 0.  \label{defgaZ}
\zee
The $\rho^{(0)}$'s are numerical constants that depend on (i) the 
coefficients of the terms in (\ref{dintVZ}) and (ii) the roots $\al_i$'s.
In the following, similar numerical factors will be denoted as 
$\rho_i^{(k)}$.  These factors are normalization factors 
and their exact values are not needed in establishing quantum 
group symmetry and calculation of the \zss-matrix.
A reminder that the fields $\zpv(x,t)$ and $\zpbv(x,t)$ are given by the
non-local expressions (\ref{nlchbZ}).
These currents obey the conservation laws 
\zbe
\zdd J^i = \zd H^i,\quad \zd \jb^i = \zdd \hb^i \quad (i=1,2,3).
\label{cuclZ}
\zee
The corresponding conserved (to lowest order) charges are
\zbe
Q^i = \frac{1}{2\pi i}\left( \int dz\, J^i + \int d\zb\, H^i\right)
\label{hahccrgZ} \zeqnalt
\zee
\zadeqn%
\zbe
\qb^i = \frac{1}{2\pi i}\left( \int d\zb\, \jb^i + \int dz\, \hb^i
\right). \zeqnalt
\zee
\zsetc%
\zeqnorg%
The Lorentz spins $s$ of the charges, defined as
$s=\Delta-\overline{\Delta}$, are
\[
s\bigl(Q^{1,2}\bigr) = s\bigl(\qb^{1,2}\bigr) = 0
\]
\zbe
s\bigl(Q^0\bigr) = -s\bigl(\qb^0\bigr) = \frac{2}{\gah}. \label{lsZ}
\zee
Note that if we
included the background charge $\zq$ (\ref{backcrgZ}), then all $Q^i$ 
(or $\qb^i$) would take on the same Lorentz spin of 
\zbe
s\bigl(Q^i\bigr) = - s\bigl(\qb^i\bigr) = \frac{1}{3}\frac{2}{\gah}, \quad i=0,1,2.
\zee
The charges are also assigned parities 
$d_i,\db_i\in\{0,1\}$, determined by the number of fermionic 
fields (i.e.\ the $\ze$'s and $\zx$'s) they contain.  
Since $Q^{1,2}$ and $\qb^{1,2}$ contain an odd number of fermions, they
are referred to as being odd or fermionic with $d_{1,2}=\db_{1,2}=1$.
The even or bosonic charges $Q^0$ and $\qb^0$, consisting of an even 
number of fermions, have $d_0=\db_0 = 0$.  All the currents have the 
same parities as the corresponding charges.  In general, the parity of
an operator $\cal{O}$ will be denoted $d(\cal{O})$.  

To obtain the algebra satisfied by the charges, we need the
braiding relations for the currents.   These can be found using
\zbe
\exp(i a \al_i\cdot\zpv(x,t) )\exp(i b \al_j\cdot\zpv(y,t) ) = 
e^{\pm i a b\pi\al_i\cdot\al_j}
\exp(i b \al_j\cdot\zpv(y,t) )\exp(i a \al_i\cdot\zpv(x,t) ), \  
x \stackrel{{\scriptstyle >}}{{\scriptstyle <}} y
\label{vbrrZ} 
\zeqnalt
\zee
\zadeqn
\zbe
\exp(i a \al_i\cdot\zpbv(x,t) )\exp(i b \al_j\cdot\zpbv(y,t) ) = 
e^{\mp i a b\pi\al_i\cdot\al_j}
\exp(i b \al_j\cdot\zpbv(y,t) )\exp(i a \al_i\cdot\zpbv(x,t) ), \  
x \stackrel{{\scriptstyle >}}{{\scriptstyle <}} y 
\zeqnalt
\zee
\zadeqn
\zbe
\exp(i a \al_i\cdot\zpv(x,t) )\exp(i b \al_j\cdot\zpbv(y,t) ) = 
e^{i a b\pi\al_i\cdot\al_j}
\exp(i b \al_j\cdot\zpbv(y,t) )\exp(i a \al_i\cdot\zpv(x,t) ), \  
\forall x,y.
\zeqnalt
\zee
\zsetc%
\zeqnorg%
To derive (\ref{vbrrZ}) one makes use of the non-local 
expressions (\ref{nlchbZ}) and the canonical commutation relations
\zbe
[\Phi_i(x,t),\zd_t\Phi_j(y,t) ] = \delta_{ij} 4 \pi i\delta(x-y).
\zee
{}From (\ref{vbrrZ}) one gets
\zbe
J^i(x,t)\jbz{j}(y,t) = (-1)^{d_i\db_j} 
e^{i\frac{\pi}{\bh^2}\al_i\cdot\al_j} \jbz{j}(y,t)J^i(x,t),
\forall x,y \label{jhbrrZ}
\zeqnalt
\zee
\zadeqn
\zbe
H^i(x,t)\hbz{j}(y,t) = (-1)^{d_i\db_j} 
e^{i\frac{\pi}{\bh^2}\al_i\cdot\al_j} \hbz{j}(y,t)H^i(x,t),
\forall x,y.
\zeqnalt
\zee
\zsetc%
\zeqnorg%
Since (\ref{jhbrrZ}b) involves applying (\ref{vbrrZ}a) and (\ref{vbrrZ}b),
its validity for all $x$ and $y$ can
be shown using a limiting procedure.  The relations (\ref{jhbrrZ}), combined 
with (\ref{gcrgbrrZ}) and (\ref{gtcrgZ}), imply that the charges satisfy
\zbe
Q^i \qbz{j} - (-1)^{d_i\db_j} e^{i\frac{\pi}{\bh^2}\al_i\cdot\al_j}
\qbz{j} Q^i =  \rho_i^{(1)}\,\delta_{ij} \frac{g \gah^2}{2\pi i}
\int_t dx\, 
\zd_x\left[ \exp\left( -i \frac{\bh}{\gah}\al_i\cdot\zPv\right)\right].
\label{crgbrrZ}
\zee
The right-hand side of (\ref{crgbrrZ}) can be
written in terms of standard topological
charges.  We take a ``soliton'' configuration
satisfying $\zPv(x=\infty)=0$,
and define the topological charges $T^i$
\zbe
T^i = \frac{\bh}{2\pi} \int_t dx\, \zd_x(\al_i\cdot \zPv ) =
- \frac{\bh}{2\pi} \al_i\cdot \zPv(x=-\infty). \label{tcrgZ}
\zee
The topological charge of any fundamental soliton field is always an integer.
Also the topological charges are not all independent.  In particular
$T^0 = -(T^1 + T^2)$, which a statement of the fact that the center is zero.
Equation (\ref{crgbrrZ}) now takes the form
\zbe
Q^i \qbz{j} - (-1)^{d_i\db_j} q_1^{-\al_i\cdot\al_j}
\qbz{j} Q^i =  \rho_i^{(1)}\,\delta_{ij} \frac{g \gah^2}{2\pi i}
(1 - q_1^{2 T^i} ), \label{crgbrr1Z}
\zee
where
\zbe 
q_1 = \exp(-i\pi/\bh^2) = -\exp(i\pi/\gah).
\zee
An equivalent expression for (\ref{crgbrrZ}) is
\zbe
\qbz{i} Q^j - (-1)^{d_j\db_i} q_2^{-\al_i\cdot\al_j}
Q^j \qbz{i} =  -\rho_i^{(1)}\,\delta_{ij} (-1)^{\db_i d_i}
q_2^{-\al_i\cdot\al_i}\,
\frac{g \gah^2}{2\pi i}
(1 - q_2^{-2 T^i} ), \label{algZ}
\zee
where $q_2$ is
\zbe 
q_2 = \frac{1}{q_1} = \exp(i\pi/\bh^2) = -\exp(-i\pi/\gah). \label{dpq2Z}
\zee
Since $\al_i\cdot\al_j$ is always an even integer, we can take
$q_i\rightarrow - q_i$ without changing (\ref{crgbrr1Z}) or (\ref{algZ}).
The braiding relations of the topological charges with the $Q$'s, which are
simply (undeformed) commutators, are found using
\zbe
\bigl[ T^i, \exp({ia\al_j\cdot\zpv + i \overline{a}\al_j\cdot\zpbv})\bigr] =
\bh (a - \overline{a}) (\al_i\cdot\al_j) 
\exp({ia\al_j\cdot\zpv + i \overline{a}\al_j\cdot\zpbv}). \label{brrtQZ}
\zee
Equation (\ref{brrtQZ}) is most easily
obtained with the complex form for $T^i$
\zbe
T^i = \frac{\bh}{2\pi}\left( \int dz\,\al_i\cdot\zd_z\zpv -
\int d\zb\, \al_i\cdot \zdb\zpbv \right). \label{cftcrgZ}
\zee
We find the commutation relations
\zbe
[T^i,Q^j] = \al_i\cdot\al_j Q^j \label{talgZ}
\zeqnalt
\zee
\zadeqn%
\zbe
[T^i,\qbz{j}] = - \al_i\cdot\al_j \qbz{j}
\zeqnalt
\zee
\zadeqn%
\zbe
[T^i,T^j] = 0.
\zeqnalt
\zee
\zsetc%
\zeqnorg%
The topological charges are even or bosonic operators, with parities
$\tilde{d}_i \equiv d(T^i) = 0$.  Note that in the above 
braiding/commutation relations, the only purpose served by
the $\ze$-$\zx$ fermion fields is to produce the correct graded
structure, namely the $(-1)^{d_i\db_j}$ factors.  Unlike the
bosons $\zp_i$ and $\zpb_i$, the fermions are treated as
being local and hence their braiding relations are simply of the
form: $\psi\overline{\psi} = - \overline{\psi}\psi$.

The equations (\ref{crgbrr1Z})/(\ref{algZ}) and (\ref{talgZ}) give the
symmetry algebra of the theory (\ref{dtacZ}) to lowest order
in perturbation theory.  In fact, this algebra is
the $q$-deformation of the the untwisted affine Lie superalgebra \ospa, 
denoted \uospa, with zero center.  
The only relations missing above are the Serre relations for \uospa. 
A review of the quantum algebra \uospa, or
quantum group as it is often called, is presented in the appendix.
Let $e_i,f_i,h_i,i=0,1,2,$ be the Chevalley generators of \uospa.  
Using the defining relations (\ref{AlsaaZ}) and (\ref{AlsaabZ}), one 
can show that the generators satisfy
\zbe
\left( e_i q^{h_i/2}\right)  \left( f_j q^{h_j/2}\right)
- (-1)^{d(e_i)d(f_j)} q^{-a_{ij}}
\left( f_j q^{h_j/2}\right)  \left( e_i q^{h_i/2}\right)
=\delta_{ij} \frac{q q^{- a_{ii}/2}}{(1-q^2)} \left(1 - q^{2 h_i}\right),
\label{chbrrZ}
\zee
where $a_{ij}\equiv \al_i\cdot\al_j$ is the  
generalized \ospa\ Cartan matrix.
Comparing (\ref{chbrrZ}) with (\ref{crgbrr1Z}) or (\ref{algZ}),
and the commutation relations (\ref{AlsaaZ}a),(\ref{AlsaaZ}b)  
with (\ref{talgZ}), we
can relate the non-local charges to the quantum group
generators in two different ways.  The first set of relations, 
which follow from considering (\ref{crgbrr1Z}) and taking $q=q_1$, are
\zbe
Q^i = c_i e_i q^{h_i/2}  \ \ (d_i = d(e_i)),\quad
\qbz{i} = c_i f_i q^{h_i/2} \ \ (\db_i = d(f_i)),  \quad
T^i = h_i \ \ (\tilde{d}_i = d(h_i)), \label{erel1Z}
\zee
where the $c_i$'s satisfy
\zbe
  c_i^2 = \rho_i^{(1)} \frac{g \gah^2}{2\pi i} q^{-1} q^{a_{ii}/2}
  (1- q^2). \label{c1Z}
\zee
The second set, obtained from (\ref{algZ}) and setting $q=q_2$, gives 
\zbe
\qbz{i} = c_i e_i q^{h_i/2} \ \ (\db_i = d(e_i)), \quad
Q^i = c_i f_i q^{h_i/2} \ \ (d_i = d(f_i)),  \quad
T^i = -h_i \ \ (\tilde{d}_i = d(h_i)),   \label{erel2Z}
\zee
with 
\zbe
  c_i^2 = -\rho_i^{(1)} (-1)^{d_i\db_i}
\frac{g \gah^2}{2\pi i} q^{-1} q^{-a_{ii}/2} (1- q^2). \label{c2Z}
\zee
(We have not yet shown $(Q^{1,2})^2 = (\qbz{{1,2}})^2 = 0$, corresponding to
the defining relation (\ref{AlsaaZ}d).  That this holds can be 
verified from the
explicit expressions in the fundamental representation discussed below.)
{}From now on we will work with the second set (\ref{erel2Z}) and take the
deformation parameter $q$ to be
\zbe
q = \exp(i\pi/\bh^2) = -\exp(-i\pi/\gah).  \label{dqZ}
\zee
Equation (\ref{dqZ}) gives the relationship between the beta parameter
$\bh$ and the quantum group deformation parameter $q$ mentioned earlier.
The marginal point ($\bh=1$) corresponds to $q=-1$.

Some important remarks regarding the relationship between the models
(\ref{todaacZ}) and (\ref{dtacZ}) and the
quantum group symmetry need to be made.  
For $\bh>1$, the model (\ref{dtacZ}) has quantum group symmetry. 
At the marginal point (\ref{dtacZ})
becomes (\ref{todaacZ}) which, when fully renormalized,
has Yangian symmetry rather than quantum group symmetry.
This means that by carefully taking the marginal limit $\bh\rightarrow 1$
of the algebra (\ref{algZ}), the Yangian symmetry algebra
can be extracted.  Or at the level of \zss-matrices, an \zss-matrix that
is symmetric under (\ref{hahccrgZ}) and (\ref{tcrgZ}) (or (\ref{erel2Z}))
should in the marginal limit give the Yangian symmetric
\zss-matrix of (\ref{seffVZ}).  So we can consider the charges
$\{Q^i,\qb^i,T^i\}$ as generating a symmetry of (\ref{dtacZ}) for
$\bh> 1$, with the understanding that for the system (\ref{seffVZ})
the limit $\bh\rightarrow 1$ needs to be taken.  The symmetry structure here
is similar to that of the affine $sl(n)$ Toda model, 
and specifically to that of the sine-Gordon model, which has $U_q(sl(2)^{(1)})$ 
symmetry for any $\bh<\sqrt{2}$ and this reduces to a Yangian symmetry
as $\bh\rightarrow \sqrt{2}$ \cite{leclairR}.  
We have obtained this symmetry algebra as a lowest order result.  Showing that
the algebra is exact to all orders in perturbation theory amounts to giving a
scaling argument that forbids any higher order terms in $g$ on the right side of
(\ref{crgbrrZ}) \cite{zamopcftR,leclairR}.  We do not discuss exactness here
since the lowest
order result is sufficient to obtain the \zss-matrix.  This is because the
constraints on $S$ placed by (i) the lowest order quantum group/Yangian
symmetry and (ii) the scattering constraints of Yang-Baxter, unitarity and
crossing, are restrictive enough so that higher order contributions
(if any) should only be of the CDD type.  In this sense the \zss-matrix
we calculate is a ``minimal'' \zss-matrix.  Lastly, in the
marginal limit it may seem that the charges (\ref{erel2Z}) blow up since
$\gah\rightarrow +\infty$.  To resolve this one can 
regularize the limit by also 
taking $g\rightarrow 0$ such that $\lim_{\bh\rightarrow 1}(g\gah)$ is
finite.  The sine-Gordon charges also need to be regularized in the marginal
limit.  Another model which displays this behavior is the 
multi-cosine model (see \cite{bogoR} and references therein).

\medskip
\noindent
{\it 3d. The fundamental fields and the comultiplication}

\smallskip
In this section we construct the quantum fields that create the particles
in the fundamental vector representation.  We also determine the 
comultiplication of the quantum charges, that is, the action of the
charges on asymptotic multiparticle states.  For completeness, we begin
with a brief review of the fundamental vector representation 
(see \cite{snrR,zhgoR,zhangR,ligoR,kacaR} for further details).

The fundamental vector representation $V$ of
\osp\ (denoted $(0,1/2)$ in \cite{snrR}) is four dimensional with a basis
$\{\zsa,\zsb,\zsc,\zsd\}$.  (We will use the same symbol for the
representation and the associated vector space.)
Since we are dealing with a Lie superalgebra,
the vector space is $Z_2$-graded and the states can be assigned parities,
$d(i)$, in two different ways: (i) $\zsa$ and $\zsd$ are even, 
$d(1)=d(4)=0$, and $\zsb$ and $\zsc$ are odd, $d(2)=d(3)=1$; or (ii)
$\zsa$ and $\zsd$ are odd, $d(1)=d(4)=1$, and $\zsb$ and $\zsc$ are even, 
$d(2)=d(3)=0$.  We take a simple root system $\{\al_1,\al_2\}$ that
is purely fermionic.  Then the Chevalley generators 
$\{e_i^V,f_i^V,h_i^V\}_{i=1,2}$
(satisfying (\ref{AlsanaZ})) in the representation $V$ are given by 
\[
e_1^V = \left( \begin{array}{cccc} 
0 & 1 & 0 & 0 \\ 0 & 0 & 0 & 0 \\ 0 & 0 & 0 & 1 \\ 0 & 0 & 0 & 0
\end{array} \right) = E_{12} + E_{34}, \quad
f_1^V = \left( \begin{array}{cccc} 
0 & 0 & 0 & 0 \\ -1 & 0 & 0 & 0 \\ 0 & 0 & 0 & 0 \\ 0 & 0 & 1 & 0
\end{array} \right) = -E_{21} + E_{43} 
\]
\[
e_2^V = \left( \begin{array}{cccc} 
0 & 0 & 1 & 0 \\ 0 & 0 & 0 & 1 \\ 0 & 0 & 0 & 0 \\ 0 & 0 & 0 & 0
\end{array} \right) = E_{13} + E_{24}, \quad
f_2^V = \left( \begin{array}{cccc} 
0 & 0 & 0 & 0 \\ 0 & 0 & 0 & 0 \\ -1 & 0 & 0 & 0 \\ 0 & 1 & 0 & 0
\end{array} \right) = -E_{31} + E_{42} 
\]
\[
h_1^V = \left( \begin{array}{cccc} 
-1 & 0 & 0 & 0 \\ 0 & -1 & 0 & 0 \\ 0 & 0 & 1 & 0 \\ 0 & 0 & 0 & 1
\end{array} \right) = -E_{11} - E_{22} + E_{33} + E_{44}
\]
\zbe
h_2^V = \left( \begin{array}{cccc} 
-1 & 0 & 0 & 0 \\ 0 & 1 & 0 & 0 \\ 0 & 0 & -1 & 0 \\ 0 & 0 & 0 & 1
\end{array} \right) = -E_{11} + E_{22} - E_{33} + E_{44}.
\label{chrepZ}
\zee
Here the $E_{ij}$'s are matrices with the only non-zero elements 
in the $i$th row and $j$th column, $(E_{ij})_{kl} = \delta_{ik}\delta_{jl}$.
The states $\zsi{i}$ can be labeled by the eigenvalues of the $h_i^V$'s
\zbe
\zsa = (-1,-1),\quad \zsb=(-1,1),\quad \zsc=(1,-1),\quad\zsd=(1,1).
\label{heigZ}
\zee
This representation can be affinized \cite{zhgobrR}, meaning that we can define
additional generators $\{e_0^V,f_0^V,h_0^V\}$ in $V$ such that
$\{e_i^V,f_i^V,h_i^V\}_{i=1,2}\cup\{x e_0^V,x^{-1}f_0^V,h_0^V\}$ 
satisfy the affine relations for \ospa\ on the loop space
$V\otimes C[x,x^{-1}]$ (with zero center).  The even simple root of the
affine extension is $\al_0=-(\al_1+\al_2)$ and the even generators are
\zbe
e_0^V = \frac{1}{\sqrt{2}} \{f_1^V, f_2^V\} = -\sqrt{2} E_{41},\quad
f_0^V = \frac{1}{\sqrt{2}} \{e_1^V, e_2^V\} = \sqrt{2} E_{14}
\label{afchZ}
\zeqnalt
\zee
\zadeqn%
\zbe
h_0^V = -(h_1^V + h_2^V) = 2 E_{11} - 2 E_{44}, 
\zeqnalt
\zee
\zsetc%
\zeqnorg%
where $\{\cdot,\cdot\}$ denotes the anticommutator.  With an abuse of
notation, the affinized representation on the loop space 
$V\otimes C[x,x^{-1}]$ will also be denoted by $V$.

Considering next the quantum algebras, the \osp\ fundamental vector
representation $V$ is undeformed as a representation of \uosp.  The
generators of \uosp\ satisfying the relations (\ref{AlsaaZ}) have 
the same matrix representations as above.  
As in the classical case, $V$ is also affinizable at the quantum level.
However for the quantum affine algebra,
the even non-Cartan generators are not given by (\ref{afchZ}a) but
are deformed to
\zbe
e_{0q}^V = - \sqrt{[2]_q} E_{41}, \quad f_{0q}^V = \sqrt{[2]_q} E_{14},
\label{afdefchZ}
\zee
where we have used the notation
\zbe
[x]_q = \frac{q^x - q^{-x}}{q- q^{-1}}. \label{bracZ}
\zee
The set 
$\{e_i^V,f_i^V,h_i^V\}_{i=1,2}\cup\{x e_{0q}^V,x^{-1}f_{0q}^V,h_0^V\}$,
forms a representation of \uospa\  on the loop space
$V\otimes C[x,x^{-1}]$.  In the limit $q\rightarrow\pm 1$, we recover
\ospa (or \osp\ from \uosp).  Note that for $q\rightarrow -1$ we have
\zbe
e_{0q}^V \rightarrow i e_0^V,\quad f_{0q}^V\rightarrow i f_0^V,
\quad [h_0]_q \rightarrow - h_0, \label{qlimZ}
\zee
and the relation (\ref{AlsaaZ}c) becomes
\zbe
[e_{0q}^V, f_{0q}^V] = [h_0]_q \stackrel{q\rightarrow -1}{\longrightarrow}
[i e_0^V, i f_0^V] = - h_0,  \label{qrellimZ}
\zee
thus recovering the classical affine relation.  Henceforth we drop
the $q$ subscript from (\ref{afdefchZ}) since
it will be either clear from the
context or explicitly stated whether we are dealing with the
quantum or classical case.

The \uosp\ representation (\ref{chrepZ}) is
referred to as being ``typical'' and type 1 or
2 grade star.  An irreducible representation $V$ is said to be typical
if any reducible representation $W$ that contains $V$ can always be
written as the direct sum $W=\tilde{W}\oplus V$, for some other
representation $\tilde{W}$.  A representation that is not typical is
said to be ``atypical''.  An atypical representation will occur in the
decomposition of $V\otimes V$ below.  The label type 1 or type 2 grade star
indicates how the representation matrices
behave under hermitian conjugation \cite{ligoR}.
A type 1 grade star representation has it generators satisfying
\[ 
(e_i^V)_{\al\bt} = (-1)^{d(f^V_i)d(\al)} \overline{(f_i^V)_{\bt\al}}
\]
\[ 
(f_i^V)_{\al\bt} = (-1)^{d(e^V_i)d(\bt)} \overline{(e_i^V)_{\bt\al}}
\]
\zbe 
(h_i^V)_{\al\bt} = \overline{(h_i^V)_{\bt\al}},
\label{t1gsrepZ}
\zee
where the overbar means complex conjugation and the subscripts $\al,\bt$
denote matrix elements.  For a type 2 grade star representation we have
\[ 
(e_i^V)_{\al\bt} = (-1)^{d(f^V_i)d(\bt)} \overline{(f_i^V)_{\bt\al}}
\]
\[ 
(f_i^V)_{\al\bt} = (-1)^{d(e^V_i)d(\al)} \overline{(e_i^V)_{\bt\al}}
\]
\zbe 
(h_i^V)_{\al\bt} = \overline{(h_i^V)_{\bt\al}}.
\label{t2gsrepZ}
\zee
If the parities are chosen to be $d(1)=d(4)=0$ and $d(2)=d(3)=0$, then
(\ref{chrepZ}) is of type 2 grade star, whereas for the second choice
$d(1)=d(4)=1$ and $d(2)=d(3)=0$, (\ref{chrepZ}) becomes type 1 grade star. 
\ignorethis{The two types of grade star representations 
lead to different hermiticity
properties for the universal $\zr$-matrix and hence to different crossing
relations for the \zss-matrix.}

We now want to construct the fields which will asymptotically create
the states/particles for the fundamental vector representation.  The
topological charges of these fields must agree with (\ref{heigZ}).
Recall that the
independent topological charges are $(T^1,T^2)=-(h_1,h_2)$.  Furthermore,
under the action of the charges the fields must transform in a manner
consistent with (\ref{chrepZ}),(\ref{afchZ}b) and (\ref{afdefchZ}).
This ensures that the states will form a representation
of the algebra (\ref{AlsaaZ}).  Specifically the charges should take the form
\zbe
\qbz{i} \propto e_i^V,\quad Q^i \propto f_i^V
\label{gaccrgZ}
\zeqnalt
\zee
\zadeqn%
\zbe
 T^i = - h_i,
\zeqnalt
\zee
\zsetc%
\zeqnorg%
when acting on the fields.

In general, the fields will consist of vertex operators multiplied by some
functions $\{f_i,\overline{f}_i\}$ of the fermions
\zbe
\pz_i = \exp\left( \frac{i}{\bh}\zw_i\cdot\zpv\right) f_i(\ze,\zd\zx),\quad
\pzb_i = \exp\left( -\frac{i}{\bh}\zw_i\cdot\zpbv\right) 
\overline{f}_i(\zeb,\zdd\zxb), \label{gfldZ}
\zee
where the $\zw_i$'s are the weights of the representation.  Both sets
$\{\pz_i\}$ and $\{\pzb_i\}$ have the same topological charges.  Any set
of fields differing from (\ref{gfldZ}) by some local operators, such as
\zbe
\chi = e^{i\al\cdot\zPv}, \label{gchifldZ}
\zee
will have the same topological charges.  This means that (\ref{gaccrgZ})
should be viewed
as modulo any local fields.  Thus knowing the topological charges, we
can look for two sets of fields, which will generate all other
``topologically equivalent'' families.  We find the following fundamental
fields giving the correct charges
\[
\pz_1(x,t) = \exp\left( \frac{i}{\bh}\zw_1\cdot\zpv(x,t)\right)\ze(z)
\]
\[
\pz_2(x,t) = \exp\left( \frac{i}{\bh}\zw_2\cdot\zpv(x,t)\right)
\]
\[
\pz_3(x,t) = \exp\left( \frac{i}{\bh}\zw_3\cdot\zpv(x,t)\right)
\]
\zbe
\pz_4(x,t) = \exp\left( \frac{i}{\bh}\zw_4\cdot\zpv(x,t)\right)\zdz\zx(z),
\label{hfldZ}
\zee
and
\[
\pzb_1(x,t) = \exp\left( \frac{i}{\bh}\zw_1\cdot\zpbv(x,t)\right)\zeb(\zb)
\]
\[
\pzb_2(x,t) = \exp\left( \frac{i}{\bh}\zw_2\cdot\zpbv(x,t)\right)
\]
\[
\pzb_3(x,t) = \exp\left( \frac{i}{\bh}\zw_3\cdot\zpbv(x,t)\right)
\]
\zbe
\pzb_4(x,t) = \exp\left( \frac{i}{\bh}\zw_4\cdot\zpbv(x,t)\right)
\zdb\zxb(\zb). \label{ahfldZ}
\zee
The weights for the fundamental representation are given by
\[ \zw_1 = (0,-i) = -\fa (\al_1+\al_2) \]
\[ \zw_2 = (1,0) = \fa (\al_1-\al_2) \]
\[ \zw_3 = (-1,0) = -\fa (\al_1-\al_2) \]
\zbe \zw_4 = (0,i) = \fa (\al_1+\al_2). \label{wgtZ} \zee
The topological charges can be found using (\ref{brrtQZ}) and
agree with (\ref{heigZ}).  Therefore
the fields (\ref{hfldZ}) and (\ref{ahfldZ}) can
be taken to produce the states
${\zsi{i}}_{i=1,\ldots,4}$ asymptotically ($t\rightarrow\pm\infty$)
\zbe
\zsi{{i,\theta}} = \pz_i \zsi{0},\quad \zsi{{i,\theta}} = \pzb_i \zsi{0},
\label{astZ}
\zee
where $\zt$ is the rapidity parameterizing the energy-momentum 
($m$ is the mass)
\zbe
E = m\cosh\theta, \quad P =m\sinh\theta. 
\zee
Since the theory (\ref{dtacZ}) is in a massive phase, the
states (\ref{astZ}) are massive
particle states with the dispersion relation $E^2 = P^2 + m^2$.  One needs
to check that the fields transform according to (\ref{gaccrgZ}).
That the topological charges are correct confirms (\ref{gaccrgZ}b).
Using the same procedure as in the previous
section, we obtain for the non-local charges acting on the fields
\[
\qbz{i}(\pz_j) = \rho_i^{(2)}\, g \gah :\sum_k (e^V_i)_{kj} \pz_k \chi_i:
\, \equiv g \sum_k (e^V_i)_{kj} \pzh_{ki}
\]
\zbe
Q^i (\pz_j) = \rho_i^{(2)}\, g \gah :\sum_k (f^V_i)_{kj} \pzb_k \chi_i:
\, \equiv g \sum_k (f^V_i)_{kj} \hat{\pzb}_{ki}, \label{crgfldacZ}
\zee
where the $\chi_i$'s are local fields
\zbe
\chi_i = \exp \left( -i \frac{\bh}{\gah} \al_i\cdot\zPv \right),
\label{chifldZ}
\zee
and $\{e_i^V,f_i^V\}_{i=0,1,2}$ are given by (\ref{chrepZ}) and (\ref{afchZ}).
{}From (\ref{crgfldacZ}) we see that (\ref{gaccrgZ}) holds.
Note that the deformed factors $\sqrt{[2]_q}$ are not obtained for
the action of $Q^0$ and $\qbz{0}$.  
This is not problematic since overall factors can be adjusted by redefining the
fields $\{\pzh_{ki},\hat{\pzb}_{ki}\}$ without changing their particle creation
properties.
Lastly, all the fields have a
non-trivial Lorentz spin.  The equations (\ref{crgfldacZ})
should have a consistent spin
structure, namely if $(e_i^V)_{kj}$ or $(f_i^V)_{kj}$ is non-zero, then
the spins must satisfy
\zbe
s(\qbz{i}) + s(\pz_j) = s(\pzh_{ki}), \quad
s(Q^i) + s(\pzb_j) = s(\hat{\pzb}_{ki}).
\zee
These relations are easily verified.  The Lorentz
spins of the charges (\ref{lsZ})
are encoded by the on-shell operators $e^{s\zt}$.  Taking into account
all the above results, the charges acting in the fundamental representation,
with states generated asymptotically by the fields (\ref{hfldZ})
and (\ref{ahfldZ}), are given by
\[
\qbz{i} = \rho_i^{(2)}\, e^{\al_i^2 \zt/ 2\gah} \,e_i^V q^{h_i^V/2}
\]
\[
Q^i = \rho_i^{(2)}\, e^{-\al_i^2\zt/2\gah}\, f_i^V q^{h_i^V/2}
\]
\zbe
 T^i = -h_i^V,  \label{crgrepZ} 
\zee
where $\{e_i^V,f_i^V,h_i^V\}_{i=0,1,2}$ are
the \uospa\ matrices (\ref{chrepZ}), (\ref{afchZ}b) and (\ref{afdefchZ}).
The Lorentz factors $\exp(\pm\al^2_i\zt/2\gah)$ play the role of the
spectral parameter $x^{\pm1}$.  Since $\al_1^2 = \al_2^2 =0$, the 
above representation corresponds to the homogeneous gradation (not to be
confused with the even/odd parity gradation).  For a non-zero background
charge $\zq$ (\ref{backcrgZ}), the $\zt$ dependence for $\qb^i$ $(Q^i),\, i=0,1,2$,
becomes $\exp(-2\zt/3\gah)$ $(\exp(+2\zt/3\gah))$.  This corresponds to the
principal gradation.  The two gradations are related by an inner automorphism.
Given an element $a(\zt) \in$ \uospa\ in the homogeneous gradation, denoted
$a_{\rm homo.}(\zt)$, the associated element in the principal gradation,
$a_{\rm princ.}(\zt)$, can be obtained from the transformation
\zbe
a_{\rm princ.}(\zt) = \sigma a_{\rm homo.}(\zt) \sigma^{-1},
\zee
where $\sigma = \exp(-h_0^V \zt/3\gah)$.

The action of the charges on multi-particle states is given by the
comultiplication.  A non-trivial comultiplication arises due to the 
non-locality of the charges and fundamental fields.  As in (\ref{jhbrrZ}), the
non-locality leads to non-trivial braiding between the currents and the 
fields.  We find the braiding relations
\[
J^i(x,t) \pzb_j(y,t) = (-1)^{d_i d(j)} q^{-T^i(\pzb_j)}
\pzb_j(y,t) J^i(x,t)
\]
\[
\jbz{i}(x,t) \pz_j(y,t) = (-1)^{d_i d(j)} q^{-T^i(\pz_j)}
\pz_j(y,t) \jbz{i}(x,t)
\]
\[
H^i(x,t) \pzb_j(y,t) = (-1)^{d_i d(j)} q^{-T^i(\pzb_j)}
\pzb_j(y,t) H^i(x,t)
\]
\zbe
\hbz{i}(x,t) \pz_j(y,t) = (-1)^{d_i d(j)} q^{-T^i(\pz_j)}
\pz_j(y,t) \hbz{i}(x,t). \label{cufldbrrZ}
\zee
Here $d(i)$ gives the parity of the fields $\pz_i$ and $\pzb_i$, and
hence of the state $\zsi{i}$: $d(i)=d(\pz_i)=d(\pzb_i)$.  (Since 
$\pz_{1,4}$ (or $\pzb_{1,4}$) contain a single fermion, we could assign
the parities $d(1)=d(4)=1$ and $d(2)=d(3)=0$.  However we will not make any
specific choice and treat both cases (i) and (ii) above consecutively.)  It
follows from (\ref{cufldbrrZ}) that the comultiplication is
\[
\zD(\qbz{i}) =  \qbz{i}\otimes 1 + q^{h_i}\otimes \qbz{i}
\]
\[
\zD(Q^i) =  Q^i\otimes 1 + q^{h_i}\otimes Q^i
\]
\zbe
\zD(h_i) =  h_i\otimes 1 + 1\otimes h_i. \label{comZ}
\zee
The expression for $\zD(h_i)$ is a consequence of the topological charge
$T^i$ being additive.  The comultiplication should be understood as
being implicitly graded, via the product definition
\zbe
(A\otimes B)(a\otimes b) = (-1)^{d(B)d(a)} A a\otimes B b, \label{grdproZ}
\zee
for charges $A,B$ and states/fields $a,b$.  This definition will be taken
to hold for all quantities with definite parities and 
where the products/actions $Aa$ and $Ba$ makes sense.  Since the 
comultiplication preserves the algebra, it provides a representation of
(\ref{erel2Z}) on multi-particle states.  By
defining a counit and an antipode, the
quantum algebra \uospa\ can be given the 
structure of a Hopf algebra \cite{zhgobrybR,brackenR,degoliyzR}.
We will only need the antipode (see section 4d), along with the 
comultiplication, to derive the \zss-matrix and hence avoid any further
discussion of this additional algebraic structure.

\addtocounter{section}{1}
\setcounter{equation}{0}

\section*{4. The \zss-matrix}

{\it 4a. Defining the \zss-matrix}

\smallskip
Given a theory with quantum affine symmetry \uga\  \cite{deliusR}, 
the one-particle states 
can be arranged into sets, or multiplets, transforming according to the
irreducible representations of \uga.  Each typical representation can be
uniquely labeled by the eigenvalues of the various Casimir operators in
the theory.  Specifically, all particles in a given multiplet have the
same mass.  Quantum group (or Yangian) symmetry ensures that there exists
an infinite set of conserved higher-spin charges in involution, implying
quantum integrability of the system.  Any scattering process is strongly
constrained by these conservation laws.  A general 
multi-particle scattering event factorizes into a series of two-particle
processes, with the set of incoming momenta and masses being equal to the
set of outgoing momenta and masses.  This means that the rapidities cannot
change and that only particles belonging to the same representation can
transform into one another.

Consider two multiplets, $\{\zsi{i}_\al\}_{i=1,\ldots,m}$ and 
$\{\zsi{j}_\bt\}_{j=1,\ldots,n}$, forming
a basis for two representations of \uga, $V_\al$ and $V_\bt$ respectively.
An asymptotic incoming two-particle state, with rapidities $\zt_1$ and
$\zt_2$, can be represented as 
\zbe
\zsi{{i,\zt_1}}_\al \otimes \zsi{{j,\zt_2}}_\bt, \label{instateZ}
\zee
where we take $\zt_1>\zt_2$ and the notation implies a spacial ordering,
i.e., the particles appear in space arranged from left to right with
decreasing rapidities.  The state (\ref{instateZ})
will scatter into some asymptotic
outgoing state which, due to the above restrictions, will be of the form
\zbe
\zsi{{k,\zt_2}}_\bt \otimes \zsi{{l,\zt_1}}_\al. \label{outstateZ}
\zee
The outgoing states will necessarily be spacially arranged from left to right
with increasing rapidities.  The two-particle \zss-matrix is an operator,
depending on the rapidities and the deformation parameter $q$ (or $\bh$),
which relates the incoming state (\ref{instateZ}) to
the outgoing state (\ref{outstateZ})
\zbe
S^{\al\bt}(\zt_1,\zt_2;q): V_\al\otimes V_\bt \longrightarrow
V_\bt \otimes V_\al  \label{smatZ}
\zee
\zbe
\zsi{{i,\zt_1}}_\al \otimes \zsi{{j,\zt_2}}_\bt =
\sum_{k,l} [S^{\al\bt}(\zt_1,\zt_2;q) ]_{ij}^{kl} \,
\zsi{{k,\zt_2}}_\bt \otimes \zsi{{l,\zt_1}}_\al.  \label{smatcompZ}
\zee
The matrix element $[S^{\al\bt}(\zt_1,\zt_2;q) ]_{ij}^{kl}$ gives the two
particle scattering amplitude for the process
\zbe
\zsi{{i,\zt_1}}_\al \otimes \zsi{{j,\zt_2}}_\bt
\longrightarrow
\zsi{{k,\zt_2}}_\bt \otimes \zsi{{l,\zt_1}}_\al. \label{sampZ}
\zee
Lorentz invariance requires that $S^{\al\bt}(\zt_1,\zt_2;q)$ only depends
on the combination $\zt\equiv\zt_1-\zt_2$
\zbe
S^{\al\bt}(\zt_1,\zt_2;q) = S^{\al\bt}(\zt;q).  \label{sliZ}
\zee
Since we are interested in the fundamental vector representation,
$V_\al = V_\bt = V$, and we will drop the $\al,\bt$ indices.

The \zss-matrix has to satisfy certain constraints.  Factorized scattering
requires that $\zs(\zt;q)$ be a solution of the Yang-Baxter 
equation \cite{yangR,baxterR,jimboybR,zzsgR,ghoshalR}.  
This fixes $\zs(\zt;q)$ up to an overall scalar constant, which can be found
by imposing the crossing and unitarity conditions 
\cite{zzsgR,ghoshalR}.  The resulting 
\zss-matrix, known as the minimal \zss-matrix, is ambiguous only up to
CDD factors.  Lastly, applying the bootstrap program fixes the CDD factors
thus giving the complete \zss-matrix \cite{ghoshalR,holloatsR,hollosbsR}.  
We will calculate the beta dependent minimal
\zss-matrix for the fundamental vector representation, leaving the bootstrap
analysis for a future problem.  The marginal \zss-matrix will then be
obtained in the limit $\bh\rightarrow 1$.

\medskip
\noindent
{\it 4b.The Yang-Baxter equation and \zrr-matrices}

\smallskip
The quantum charges (\ref{erel2Z}) generate a symmetry of the theory.
Therefore the action of the \zss-matrix must commute with the action, or
comultiplication, of the charges
\zbe
[S,\zD(\qbz{i}) ] = [S,\zD(Q^i)] = [S,\zD(h_i)] = 0. \label{screlZ}
\zee
For the fundamental representation (\ref{crgrepZ}), these commutation
relations take the explicit form
\zbe
S(\zt;q) \left( x_{1i} e_i^V q^{h_i^V/2}\otimes 1 +
q^{h_i^V} \otimes x_{2i} e_i^V q^{h_i^V/2} \right) =
\left( x_{2i} e_i^V q^{h_i^V/2}\otimes 1 +
q^{h_i^V} \otimes x_{1i} e_i^V q^{h_i^V/2} \right) S(\zt;q) 
\label{sitZ}
\zeqnalt
\zee
\zadeqn%
\zbe
S(\zt;q) \left( x_{1i}^{-1} f_i^V q^{h_i^V/2}\otimes 1 +
q^{h_i^V} \otimes x_{2i}^{-1} f_i^V q^{h_i^V/2} \right) =
\left( x_{2i}^{-1} f_i^V q^{h_i^V/2}\otimes 1 +
q^{h_i^V} \otimes x_{1i}^{-1} f_i^V q^{h_i^V/2} \right) S(\zt;q) 
\zeqnalt
\zee
\zadeqn%
\zbe
S(\zt;q)\left( h_i^V\otimes 1 + 1\otimes h_i^V\right) =
\left( h_i^V\otimes 1 + 1\otimes h_i^V\right) S(\zt;q),
\zeqnalt
\zee
\zsetc%
\zeqnorg%
where
\zbe
x_{ji} = e^{\al_i^2 \zt_j/2\gah}, \ \ i=0,1,2;\ j=1,2.
\label{defxji}
\zee
For the fermionic roots we simply have $x_{j1}=x_{j2}=1$.  Since 
$h_0^V=-(h_1^V + h_2^V)$, the equation involving $\zD(h_0^V)$ need not be
considered independently.  Multiplying both sides of (\ref{sitZ}) by
$q^{-h_i^V/2}\otimes q^{-h_i^V/2}$ from the right, and making use of
\zbe
S(\zt;q) \left( q^{-h_i^V/2}\otimes q^{-h_i^V/2}\right)
= \left( q^{-h_i^V/2}\otimes q^{-h_i^V/2}\right)S(\zt;q),  \label{scomhZ}
\zee
which follows from (\ref{sitZ}c), the expressions (\ref{sitZ}) can 
be rewritten as
\zbe
[S(x;q),\zD(e^V_{1,2})] = 0
\label{scomrelZ}
\zeqnalt
\zee
\zadeqn%
\zbe
[S(x;q),\zD(e^V_0)] = 0
\zeqnalt
\zee
\zadeqn%
\zbe
[S(x;q),\zD(f^V_{1,2})] = 0
\zeqnalt
\zee
\zadeqn%
\zbe
[S(x;q),\zD(f^V_0)] = 0
\zeqnalt
\zee
\zadeqn%
\zbe
[S(x;q),\zD(h^V_{1,2})] = 0, 
\zeqnalt
\zee
\zsetc%
\zeqnorg%
where $\zD(h_{1,2}^V)$ is as in (\ref{comZ}) and
\[
\zD(e_{1,2}^V) = e^V_{1,2}\otimes q^{-h^V_{1,2}/2} + 
q^{h_{1,2}^V/2}\otimes e^V_{1,2}
\]
\[
\zD(e_{0}^V) = x_{10} e^V_{0}\otimes q^{-h^V_{0}/2} + 
q^{h_{0}^V/2}\otimes x_{20} e^V_{0}
\]
\[
\zD(f_{1,2}^V) = f^V_{1,2}\otimes q^{-h^V_{1,2}/2} + 
q^{h_{1,2}^V/2}\otimes f^V_{1,2}
\]
\zbe
\zD(f_{0}^V) = x_{10}^{-1} f^V_{0}\otimes q^{-h^V_{0}/2} + 
q^{h_{0}^V/2}\otimes x_{20}^{-1} f^V_{0}. \label{comefZ}
\zee
The comultiplication (\ref{comefZ}) is the standard comultiplication for the
Chevalley generators of \uospa.  For the affine generators, $e_0$ and $f_0$,
the comultiplication corresponds to the representation $e_0 = x_{i0} e_0^V$
and $f_0 = x_{i0}^{-1} f_0^V$ over the loop algebra.
We have also defined the spectral parameter
\zbe
x = \frac{x_{10}}{x_{20}} = e^{-2(\zt_1-\zt_2)/\gah} = e^{-2\zt/\gah},
\label{spdefZ}
\zee
and indicated the dependence of \zss\ on $\zt$ implicitly via $x$.

Equation (\ref{scomrelZ}) is precisely one of
the defining relations for the \uospa\
\zrr-matrix $\rt(x;q)$ in the homogeneous gradation.  More generally, for
affinizable representations $V_\al$ and $V_\bt$, 
$\rt^{\al\bt}(x)$ acts as $S(x)$
\zbe
\rt(x;q): V_\al\otimes V_\bt \longrightarrow V_\bt\otimes V_\al, \label{acRZ}
\zee
and satisfies the intertwining property \cite{degoliyzR}
\[
\rt^{\al\bt}(x,q) \zD^{\al\bt}(a) =
\zD^{\bt\al}(a) \rt^{\al\bt}(x,q), \ \ \forall a\in 
U_q[osp(2|2)]
\]
\[
\rt^{\al\bt}(x;q) \left( x e_0^{\al} \otimes q^{-h_0^{\bt}/2} +
q^{h_0^{\al}/2} \otimes e_0^{\bt} \right) =
\left( e_0^{\bt} \otimes q^{-h_0^{\al}/2} +
q^{h_0^{\bt}/2} \otimes x e_i^{\al}  \right) \rt^{\al\bt}(x;q) 
\]
\zbe
\rt^{\al\bt}(x;q) \left( x^{-1} f_0^{\al} \otimes q^{-h_0^{\bt}/2} +
q^{h_0^{\al}/2} \otimes f_0^{\bt} \right) =
\left( f_0^{\bt} \otimes q^{-h_0^{\al}/2} +
q^{h_0^{\bt}/2} \otimes x^{-1} f_i^{\al}  \right) \rt^{\al\bt}(x;q),
\label{ritZ}
\zee
where all quantities are evaluated in the appropriate representation as  
indicated by the $\al,\bt$ indices.  One can also view $\rt^{\al\bt}(x)$ as
the spectral parameter dependent \zrr-matrix of \uosp.  This equivalence
follows from the fact that for affinizable representations, both the
\uospa\ \zrr-matrix (in the homogeneous gradation) and the spectral
parameter dependent \zrr-matrix of \uosp\ satisfy the same defining relation
(\ref{ritZ}).  Another \zrr-matrix $R(x;q)$, commonly referred to as the
universal \zrr-matrix, can be obtained from $\rt(x;q)$ by applying a
permutation operation.  Define the graded permutation operator $P$ satisfying
\zbe
P(\zsi{u}\otimes\zsi{v}) = (-1)^{d(u)d(v)} \zsi{v}\otimes \zsi{u},
\label{permdefz}
\zee
for any states $\zsi{u}$ and $\zsi{v}$ in $V$ with definite parity.  The
universal \zrr-matrix is related to $\rt(x;q)$ by
\zbe
R(x;q) = P \rt(x;q). \label{urdefZ}
\zee
The universal \zrr-matrix is known for various quantum affine superalgebras
(see \cite{degoliyzR} and references therein).

For bosonic quantum groups, Jimbo \cite{jimboR} showed 
that a solution to (\ref{ritZ})
is unique up to an overall scalar function, and that any solution 
automatically satisfies the Yang-Baxter equation.  These results were 
extended to quantum supergroups by Bracken {\it et al.} \cite{brackenR}.  
These authors showed
that a non-trivial solution of (\ref{scomrelZ}a) and (\ref{scomrelZ}b)
must be even and is unique up to scalar factors.
Furthermore, this solution satisfies (i) equations
(\ref{scomrelZ}c)--(\ref{scomrelZ}e) and (ii) the Yang-Baxter equation.
The uniqueness of the solution implies that $S(x;q)$ takes the form
\zbe
S(x;q) = v(x;q) \rt(x;q),  \label{sformZ}
\zee
where the scalar function $v(x;q)$ is fixed by imposing the 
unitarity and crossing constraints (see section 4e) and by applying the
bootstrap principle.  Thus for a theory with quantum group symmetry, the
Yang-Baxter equation need not be independently solved.  Rather factorization
is a consequence of the intertwining property (\ref{ritZ}) of the
\zrr-matrix.

The Yang-Baxter equations satisfied by the two \zrr-matrices are
\zbe
R_{12}(x;q) R_{13}(xy;q)R_{23}(y;q) = R_{23}(y;q)R_{13}(xy;q) R_{12}(x;q)
\label{yburZ}
\zee
\zbe
\rt_{23}(x;q)\rt_{12}(xy;q)\rt_{23}(y;q) =
\rt_{12}(y;q)\rt_{23}(xy;q)\rt_{12}(x;q).
\label{ybrZ}
\zee
Both equations act on $V\otimes V\otimes V$ and for a solution
$R=\sum_i a_i\otimes b_i$, the notation $R_{ij}$ means
\zbe
R_{12} = \sum_i a_i\otimes b_i \otimes 1,\ \ 
R_{13} = \sum_i a_i\otimes 1 \otimes b_i, \ \ 
R_{23} = \sum_i 1\otimes a_i\otimes b_i,  \label{runotZ}
\zee
and similarly for $\rt_{ij}$.  The Yang-Baxter equation for $R(x;q)$ 
is implicitly graded due to the multiplication rule (\ref{grdproZ}).
In components (\ref{yburZ}) takes the form (summing over repeated indices)
\begin{eqnarray}
\lefteqn{R(x;q)^{b_1 b_2}_{c_1 c_2} R(xy;q)^{c_1 b_3}_{a_1 c_3} 
R(y;q)^{c_2 c_3}_{a_2 a_3} (-1)^{d(b_1)d(b_2) + d(c_1)d(b_3)+ d(c_2)d(c_3)}
 =} \nonumber \\ & & \quad\quad\quad\quad\quad\quad
R(y;q)^{b_2 b_3}_{c_2 c_3}R(xy;q)^{b_1 c_3}_{c_1 a_3} 
R(x;q)^{c_1 c_2}_{a_1 a_2} (-1)^{d(b_2)d(b_3) + d(b_1)d(c_3)+ d(c_1)d(c_2)}.
\label{yburcompZ}
\end{eqnarray}
The \zrr-matrix 
$\tilde{R}(x;q)_{a_1 a_2}^{b_1 b_2} \equiv R(x;q)_{a_1 a_2}^{b_1 b_2} 
(-1)^{d(b_1)d(b_2)}$ satisfies the ordinary Yang-Baxter equation.  The 
component form of (\ref{ybrZ}) can be obtained from (\ref{yburcompZ})
by setting
\zbe
\rt(x;q)_{a_1 a_2}^{b_1 b_2} = R(x;q)_{a_1 a_2}^{b_2 b_1}
(-1)^{d(b_1)d(b_2)}, \label{rurrelZ}
\zee
which does not contain any parity factors.  Of course the \zss-matrix
also satisfies (\ref{ybrZ}).

\medskip
\noindent
{\it 4c. The \uospa\  \zrr-matrix $\rt(x;q)$}

\smallskip
A solution to (\ref{scomrelZ})/(\ref{ritZ}) has been previously
computed in \cite{massaR,goliyztsR} and \cite{deguchiR}.   
In \cite{massaR} $\rt(x;q)$ was
constructed for a set of four-dimensional typical
representations characterized by a continuous parameter $b$.  This set
includes the fundamental vector representation.  We will re-derive $\rt(x;q)$
using the method of \cite{goliyztsR} and including some of the details
omitted in these references.  Our conventions are also slightly different.
We hope a more explicit construction of $\rt(x;q)$, showing the various
steps involved, will be useful.

As discussed above, it is sufficient to solve the following reduced set
of equations
\[
[\rt(x;q),\zD(e_{1,2}^V) ] = 0
\]
\zbe
\rt(x;q) \left( x e_0^V \otimes q^{-h_0^V/2} + q^{h_0/2}\otimes e_0^V\right)
=\left( e_0^V \otimes q^{-h_0^V/2} + q^{h_0/2}\otimes x e_0^V\right)
\rt(x;q). \label{reditZ}
\zee
However it is equally as manageable, and perhaps more illustrative, to
work with the set
\zbe
[\rt(x;q),\zD(a) ] = 0,\  \forall a\in U_q[osp(2|2)]
\label{rediteZ}
\zeqnalt
\zee
\zadeqn%
\zbe
\rt(x;q) \left( x e_0^V \otimes q^{-h_0^V/2} + q^{h_0/2}\otimes e_0^V\right)
=\left( e_0^V \otimes q^{-h_0^V/2} + q^{h_0/2}\otimes x e_0^V\right)
\rt(x;q),
\zeqnalt
\zee
\zsetc%
\zeqnorg%
which obviously includes (\ref{reditZ}).
Thus we will seek a solution to (\ref{rediteZ}).

Consider for a moment the general case (\ref{ritZ}), with (\ref{rediteZ})
being evaluated in $V_\al\otimes V_\bt$ for affinizable representations
$V_\al$ and $V_\bt$.  Suppose $V_\al\otimes V_\bt$ has the
multiplicity-free tensor product decomposition
\zbe
V_\al \otimes V_\bt = \bigoplus_\mu V_\mu, \label{gtpdZ}
\zee
into \uosp\ invariant spaces $V_\mu$.  Since $\rt(x;q)$ commutes with
the \uosp\ comultiplication, a solution of (\ref{rediteZ}a) can be written as
\zbe
\rt(x;q) = \sum_\mu \rho_\mu(x;q) \cp_\mu^{\al\bt}(q), \label{rsolaZ}
\zee
where $\cp_\mu^{\al\bt}(q)$ are projectors onto $V_\mu$ and $\rho_\mu(x;q)$
are arbitrary functions.  If the decomposition (\ref{gtpdZ}) is
completely reducible, then (\ref{rsolaZ}) is the most
general solution of (\ref{rediteZ}a).  If instead the decomposition
is not fully reducible, then (\ref{rsolaZ}) need not be the most
general solution, though it certainly is one solution.

Returning to the fundamental representation, we look for a decomposition
of $V\otimes V$ into \uosp\ invariant spaces.  In the classical case, the
tensor product is not completely reducible.  One finds a decomposition
into two eight-dimensional invariant spaces \cite{snrR}
\zbe
V\otimes V = W \oplus \wt,  \label{tpdZ}
\zee
where $W$ is an irreducible \osp\ representation spanned by
\begin{eqnarray}
\zsi{\chi_1^\pm} & = & \ztp{1}{1} \nonumber \\
\zsi{\chi_2^\pm} & = & \fb ( \ztp{1}{2}\pm \ztp{2}{1} ) \nonumber \\ 
\zsi{\chi_3^\pm} & = & \fb ( \ztp{1}{3}\pm \ztp{3}{1} ) \nonumber \\ 
\zsi{\chi_4^\pm} & = & \fb ( \ztp{2}{3} - \ztp{3}{2} ) \nonumber \\ 
\zsi{\chi_5^\pm} & = & \fb (\ztp{1}{4} + \ztp{4}{1} )\nonumber \\ 
\zsi{\chi_6^\pm} & = & \fb (\ztp{2}{4} \pm \ztp{4}{2} ) \nonumber \\ 
\zsi{\chi_7^\pm} & = & \fb (\ztp{3}{4} \pm \ztp{4}{3} ) \nonumber \\ 
\zsi{\chi_8^\pm} & = & \ztp{4}{4},  \label{clst1Z}
\end{eqnarray}
and  $\wt$  is an atypical representation spanned by
\begin{eqnarray}
\zsi{\chi_9^\pm} & = & \ztp{2}{2} \nonumber \\
\zsi{\chi_{10}^\pm} & = & \fb ( \ztp{1}{2}\mp \ztp{2}{1} ) \nonumber \\ 
\zsi{\chi_{11}^\pm} & = & \fb ( \ztp{1}{3}\mp \ztp{3}{1} ) \nonumber \\ 
\zsi{\chi_{12}^\pm} & = & \fa ( \ztp{1}{4} - \ztp{4}{1} \mp \ztp{2}{3} 
\mp \ztp{3}{2} ) \nonumber \\ 
\zsi{\chi_{13}^\pm} & = & \fa (\ztp{1}{4} - \ztp{4}{1} \pm \ztp{2}{3}
\pm \ztp{3}{2} )\nonumber \\ 
\zsi{\chi_{14}^\pm} & = & \fb (\ztp{2}{4} \mp \ztp{4}{2} ) \nonumber \\ 
\zsi{\chi_{15}^\pm} & = & \fb (\ztp{3}{4} \mp \ztp{4}{3} ) \nonumber \\ 
\zsi{\chi_{16}^\pm} & = & \ztp{3}{3}.  \label{clst2Z}
\end{eqnarray}
Here and henceforth the upper (lower) sign is to be taken if the parities
are chosen to be even (odd) for $\zsi{1}$, $\zsi{4}$ and odd (even) for
$\zsi{2}$, $\zsi{3}$, and the associated states will be 
labeled by a + ($-$) superscript.  The two sets of states 
\zbe
\{\zsi{\chi_{9}^\pm},\zsi{\chi_{10}^\pm},
\zsi{\chi_{11}^\pm},\zsi{\chi_{12}^\pm} \}\qquad
{\rm and} \qquad
\{\zsi{\chi_{12}^\pm},\zsi{\chi_{14}^\pm},
\zsi{\chi_{15}^\pm},\zsi{\chi_{16}^\pm} \}, \label{atysZ}
\zee
form four-dimensional atypical representations, with $\zsi{\chi_{12}^\pm}$
being invariant, i.e., is mapped to zero by all the generators.  The
state $\zsi{\chi_{13}^\pm}$ is a cyclic vector for $\wt$.  In the quantum 
case these states will be deformed, yet one expects that the basic structure
of the decomposition will be the same, namely the tensor product is not
completely reducible and has the form \cite{massaR,goliyztsR}
\zbe
V\otimes V = W_q \oplus \wt_q,  \label{qtpdZ}
\zee
where $W_q$ is an irreducible invariant space going over to $W$ as
$q\rightarrow 1$, and $\wt_q$ is not irreducible, but contains an invariant
singlet state and goes over to $\wt$ as $q\rightarrow 1$.  A mechanical way
to determine a basis for, say $W_q$, is to start with one of the states
$\zsi{\chi_{i}^\pm}\in W$, or some deformed version of it.  Acting on this 
state with the \uosp\ generators will result in some new deformed states.
One keeps repeating the process, acting on every new state with
$\zD(a)$ until an
invariant set with the required properties is obtained.  The success of this
procedure depends on the choice of the initial state.  Starting with the
undeformed states $\ztp{1}{1}$, $\ztp{2}{2}$ and $\ztp{3}{3}$, almost
all the basis states of $W_q$ and $\wt_q$ can be obtained this way.  The 
exception is the cyclic state of $\wt_q$ which has to be independently 
constructed.  We find the following set spanning $W_q$
\begin{eqnarray}
\zsi{\psi^\pm_{1}} & = & \ztp{1}{1} \nonumber \\
\zsi{\psi^\pm_{2}} & = & \fc\left( q^{-1/2}\ztp{1}{2}\pm q^{1/2}\ztp{2}{1} 
\right)\nonumber \\
\zsi{\psi^\pm_{3}} & = & \fc\left( q^{-1/2}\ztp{1}{3}\pm q^{1/2}\ztp{3}{1} 
\right)\nonumber \\
\zsi{\psi^\pm_{4}} & = & \fd\left( q^{-1}\ztp{1}{4}+ q\ztp{4}{1} \pm 
q\ztp{2}{3}\mp q^{-1}\ztp{3}{2} \right)\nonumber \\
\zsi{\psi^\pm_{5}} & = & \fd\left( q^{-1}\ztp{1}{4}+ q\ztp{4}{1} \mp 
q^{-1}\ztp{2}{3}\pm q\ztp{3}{2} \right)\nonumber \\
\zsi{\psi^\pm_{6}} & = & \fc\left( q^{-1/2}\ztp{2}{4}\pm q^{1/2}\ztp{4}{2} 
\right)\nonumber \\
\zsi{\psi^\pm_{7}} & = & \fc\left( q^{-1/2}\ztp{3}{4}\pm q^{1/2}\ztp{4}{3} 
\right)\nonumber \\
\zsi{\psi^\pm_{8}} & = & \ztp{4}{4}, \label{qsttyZ}
\end{eqnarray}
and for $\wt_q$ a basis is given by
\begin{eqnarray}
\zsi{\psi^\pm_{9}} & = & \ztp{2}{2} \nonumber \\
\zsi{\psi^\pm_{10}} & = & \fc\left( q^{1/2}\ztp{1}{2}\mp q^{-1/2}\ztp{2}{1} 
\right)\nonumber \\
\zsi{\psi^\pm_{11}} & = & \fc\left( q^{1/2}\ztp{1}{3}\mp q^{-1/2}\ztp{3}{1} 
\right)\nonumber \\
\zsi{\psi^\pm_{12}} & = & \fa\left( \ztp{1}{4}-\ztp{4}{1} \mp
\ztp{2}{3}\mp \ztp{3}{2} \right)\nonumber \\
\zsi{\psi^\pm_{13}} & = & \frac{1}{\sqrt{q^2 + q^{-2}}}
\left( q \ztp{1}{4} - q^{-1}\ztp{4}{1} \right)\nonumber \\
\zsi{\psi^\pm_{14}} & = & \fc\left( q^{1/2}\ztp{2}{4}\mp q^{-1/2}\ztp{4}{2} 
\right)\nonumber \\
\zsi{\psi^\pm_{15}} & = & \fc\left( q^{1/2}\ztp{3}{4}\mp q^{-1/2}\ztp{4}{3} 
\right)\nonumber \\
\zsi{\psi^\pm_{16}} & = & \ztp{3}{3}. \label{qstatyZ}
\end{eqnarray}
One can check that: \newline
(i)
\zbe
\lim_{q\rightarrow 1} W_q = \lim_{q\rightarrow 1}\,{\rm span}\{
\zsi{\psi^\pm_i};i=1,\ldots,8\} = W,\qquad
\lim_{q\rightarrow 1} \wt_q = \lim_{q\rightarrow 1}\,{\rm span}\{
\zsi{\psi^\pm_i};i=9,\ldots,16\} = \wt;
\zee
(ii) the set (\ref{qsttyZ}) forms a typical
representation of \uosp; (iii) the space
$\wt_q$ is composed of two atypical \uosp\ representations spanned by
\zbe
\{\zsi{\psi_{9}^\pm},\zsi{\psi_{10}^\pm},
\zsi{\psi_{11}^\pm},\zsi{\psi_{12}^\pm} \}\qquad
{\rm and} \qquad
\{\zsi{\psi_{12}^\pm},\zsi{\psi_{14}^\pm},
\zsi{\psi_{15}^\pm},\zsi{\psi_{16}^\pm} \}; \label{atrepsZ}
\zee
(iv) the state $\zsi{\psi^\pm_{12}}$ is a singlet state which cannot be
separated from the atypical representations; and (v) $\zsi{\psi^\pm_{13}}$
is a cyclic vector for $\wt_q$ which cannot be obtained by acting on any
basis state by any of the generators.  The adjoint states $\zsj{\psi_i^\pm}$
are given by (note we do not take $q\rightarrow \overline{q}$ for the
adjoint states)
\zbe
\zsj{\psi_i^\pm} = \left(\zsi{\psi_i^\pm} \right)^\dagger, \label{adstZ}
\zee
where
\zbe
\left( \ztp{i}{j}\right)^\dagger = (-1)^{d(i)d(j)} \zsj{i}\otimes\zsj{j} 
\label{addefZ}
\zee
\zbe
\left( \zsi{i}\right)^\dagger = \zsj{i}, \ \ i=1,2,3,4.
\label{sadZ}
\zee
The parity factor in (\ref{addefZ}) is introduced to cancel that
of (\ref{grdproZ}), so that the norm is positive
\zbe
\left( \ztp{i}{j}\right)^\dagger \left( \ztp{i}{j}\right) =
\langle i | i\rangle\langle j | j \rangle = 1. \label{posnormZ}
\zee

To determine $\rt(x;q)$ we need to know the projectors for $W_q$ and
$\wt_q$, denoted $\cp_1(q)$ and $\cp_0(q)$ respectively.  If the states were
all orthonormal, then $\cp_1(q)$ and $\cp_0(q)$ would have the usual form:
$\sum_i \zsi{\psi^\pm_i}\zsj{\psi^\pm_i}$.  However the four states
$\{\zsi{\psi^\pm_4},\zsi{\psi^\pm_5},\zsi{\psi^\pm_{12}},
\zsi{\psi^\pm_{13}}\}$ are not orthonormal.  Following \cite{goliyztsR}, 
we define the dual states $\{ \zsj{\hat{\psi}_i^\pm}\}$ satisfying
\zbe
\langle \hat{\psi}^\pm_i | \psi^\pm_j\rangle = \delta_{ij}. \label{dstZ}
\zee
Let $g_{ij}$ be the metric
\zbe
g_{ij} = \langle \psi^\pm_i | \psi^\pm_j\rangle, \label{gmetZ}
\zee
then the dual states are given by
\zbe
\zsj{\hat{\psi}^\pm_i} = \sum_j (g^{-1})_{ij} \zsj{\psi^\pm_j}.
\label{dstdefZ}
\zee
For $i \notin \{4,5,12,13\}$, we simply have
\zbe
\zsj{\hat{\psi}^\pm_i} = \zsj{\psi^\pm_i}. \label{usdstZ}
\zee
Using the dual states the projectors can be written as
\zbe
\cp_1(q) = \sum_{i=1}^8 \zsi{\psi_i^\pm} \zsj{\hat{\psi}_i^\pm}
\label{proj1Z}
\zee
\zbe
\cp_0(q) = \sum_{i=9}^{16} \zsi{\psi_i^\pm} \zsj{\hat{\psi}_i^\pm}.
\label{proj0Z}
\zee
It is easily shown that (\ref{proj1Z}) and (\ref{proj0Z}) obey
\zbe
\cp_1(q)^2 = \cp_1(q),\quad \cp_0(q)^2 = \cp_0(q),
\quad \cp_1(q)\cp_0(q) = \cp_0(q)\cp_1(q) = 0,
\quad \cp_1(q) + \cp_0(q) = 1. \label{propprojZ}
\zee
A particular solution for $\rt(x;q)$  satisfying (\ref{rediteZ}a) is 
now given by
\zbe
\rt_0(x;q) = \rho_1(x;q) \cp_1(q) + \rho_0(x;q)\cp_0(q). \label{partsrZ}
\zee

Recalling our previous comments, since the decomposition (\ref{qtpdZ}) is not
completely reducible, (\ref{partsrZ}) is not necessarily the most
general solution. In particular, the operator
\zbe
\cp_N(q) = \zsi{\psi^\pm_{12}}\zsj{\hat{\psi}^\pm_{13}}, \label{projNZ}
\zee
mapping the cyclic vector onto the singlet state, can be
added to (\ref{partsrZ}).
That this does not spoil (\ref{rediteZ}a) follows 
from (iv) and (v) above, which
imply
\zbe
\Delta(a) \cp_N(q) \zsi{\psi^\pm_i} = \cp_N(q) 
\Delta(a) \zsi{\psi^\pm_i} = 0, \ \ \forall i.
\zee
Some useful properties of $\cp_N(q)$ are
\zbe
\cp_N(q)^2 = 0,\quad \cp_1(q)\cp_N(q) = \cp_N(q)\cp_1(q) = 0,\quad
\cp_0(q)\cp_N(q) = \cp_N(q)\cp_0(q) = \cp_N(q),
\label{proppNZ}
\zee
with the first expressing the order 2 nilpotency of $\cp_N(q)$.  Thus the
most general solution of (\ref{rediteZ}a) is
\zbe
\rt(x;q) = \rho_1(x;q) \cp_1(q) + \rho_0(x;q)\cp_0(q) + \rho_N(x;q)\cp_N(q),
\label{gensolrZ}
\zee
where the $\rho_i(x;q)$'s are at present arbitrary functions soon to be
fixed by imposing (\ref{rediteZ}b).

Before calculating these functions, we give the explicit expressions
for the projectors and $\cp_N(q)$
in the original basis $\{\ztp{i}{j};i,j=1,\ldots,4\}$.
We choose to order the basis as
\zbe
(v_{11},v_{22},v_{33},v_{44},v_{12},v_{21},v_{13},v_{31},v_{24},v_{42},
v_{34},v_{43},v_{14},v_{23},v_{32},v_{41}),
\label{basisorZ}
\zee
where
\zbe
v_{ij} \equiv \ztp{i}{j}. \label{vijdefZ}
\zee
Define $\hat{v}$ as the column vector associated with (\ref{basisorZ}), i.e., 
$\hat{v}_1=v_{11},\hat{v}_2=v_{22},\ldots$.  The basis states can be written 
as
\zbe
\zsi{\psi^\pm_i} = \sum_j M_{ij} \hat{v}_j, \label{bstMZ}
\zee
where the matrix $M_{ij}$ can be obtained from (\ref{qsttyZ})
and (\ref{qstatyZ}).  The metric and the inverse metric take
the form ($T$ denotes transpose)
\zbe
g = M M^T,\qquad g^{-1} = (M^{-1})^T M^{-1}. \label{gmetMZ}
\zee
This gives the single ``projectors''
\[
\cp_{kl}(q) \equiv \zsi{\psi^\pm_k} \zsj{\hat{\psi}^\pm_l}
\]
\zbe
(\cp_{kl}(q))_{ij} = M_{ki} (M^{-1})_{jl}, \label{sglprojZ}
\zee
and hence for the required matrix components 
\zbe
(\cp_1(x;q))_{ij} = \sum_{k=1}^8 M_{ki} (M^{-1})_{jk} \label{proj1MZ}
\zee
\zbe
(\cp_0(x;q))_{ij} = \sum_{k=9}^{16} M_{ki} (M^{-1})_{jk} \label{proj0MZ}
\zee
\zbe
(\cp_N(x;q))_{ij} =  M_{12i} (M^{-1})_{j13}. \label{projNMZ}
\zee
These expressions are easily evaluated.  We find the following block diagonal
form for all the operators ($\zo = \cp_{1,0,N}(x;q)$)
\zbe
\zo_{16\times 16} =
\left( \begin{array}{cccccccccc}
\zo_{1\times 1} & & & & & & & &  \\
& \zo_{1\times 1} & & & & & & &  \\
& & \zo_{1\times 1} & & & & & &  \\
& & & \zo_{1\times 1} & & & & &  \\
& & & & \zo_{2\times 2} & & & &  \\
& & & & & \zo_{2\times 2} & & &  \\
& & & & & & \zo_{2\times 2} & &  \\
& & & & & & & \zo_{2\times 2} &  \\
& & & & & & & & \zo_{4\times 4} 
\end{array} \right),
\label{bldfZ}
\zee
with the individual blocks being: \newline
(i) One-dimensional blocks:
\zbe
\cp_1(q) = 1,\ \ \cp_0(q) = \cp_N(q) = 0,
\zee
for  $\ztp{1}{1}$ and $\ztp{4}{4}$; and
\zbe
\cp_0(q) = 1, \ \ \cp_1(q) = \cp_N(q) = 0,
\zee
for $\ztp{2}{2}$ and $\ztp{3}{3}$. \newline
(ii) Two-dimensional blocks:
\zbe
\cp_1(q) = \frac{1}{q^2 + 1} \left( 
\begin{array}{cc}
1 & \pm q \\
\pm q & q^2 
\end{array}
\right), \ \ 
\cp_0(q) = \frac{1}{q^2 + 1} \left( 
\begin{array}{cc}
q^2 & \mp q \\
\mp q & 1 
\end{array}
\right), \ \ 
\cp_N(q) = 0,
\zee
for the four pairs of states 
\[
(\ztp{1}{2},\ztp{2}{1}), \ \
(\ztp{1}{3},\ztp{3}{1}), \ \ 
(\ztp{2}{4},\ztp{4}{2}), \ \ 
(\ztp{3}{4},\ztp{4}{3}).
\]
(ii) Four-dimensional blocks:
\zbe
\cp_1(q) = \frac{1}{(q^2 + 1)^2} \left(
\begin{array}{cccc}
2  & \mp(q^2 - 1) & \mp(q^2 - 1) & 2 q^2 \\
\pm(q^2 -1) & 2 q^2 & -q^4 - 1 & \pm(q^4 - q^2) \\
\pm(q^2 -1) & -q^4 - 1 & 2 q^2 & \pm(q^4 - q^2) \\
2 q^2 & \mp(q^4 - q^2) & \mp(q^4-q^2) & 2 q^4 
\end{array}
\right),
\zee
\zbe
\cp_0(q) = \frac{1}{(q^2 + 1)^2} \left(
\begin{array}{cccc}
q^4 + 2 q^2 - 1  & \pm(q^2 - 1) & \pm(q^2 - 1) & -2 q^2 \\
\mp(q^2 -1) & q^4 + 1 & q^4 + 1 & \mp(q^4 - q^2) \\
\mp(q^2 -1) & q^4 + 1 & q^4 + 1 & \mp(q^4 - q^2) \\
-2 q^2 & \pm(q^4 - q^2) & \pm(q^4-q^2) & -q^4 +2q^2+1
\end{array}
\right),
\zee
\zbe
\cp_N(q) = \frac{\sqrt{q^4 + 1}}{2(q^2 + 1)} \left(
\begin{array}{cccc}
1 & \pm 1 & \pm 1 & -1 \\
\mp 1 & -1 & -1 & \pm 1\\
\mp 1 & -1 & -1 &\pm 1\\
-1 & \mp 1 & \mp 1 & 1
\end{array}
\right),
\zee
for the basis states
\[
(\ztp{1}{4},\ztp{2}{3},\ztp{3}{2},\ztp{4}{1}).
\]
The projectors for the two different parity assignments, denote them
$\cp_i^+(q)$ and $\cp_i^-(q)$, are related by a similarity transformation
\zbe
\cp_i^+(q) = G \cp_i^-(q) G,\qquad G^2 = 1,
\label{simtransrelZ}
\zee
where $G$ is the diagonal matrix
\zbe
G = {\rm diag}(1,1,1,1,1,-1,1,-1,1,-1,1,-1,1,-1,-1,1). \label{simtransZ}
\zee
Obviously, the corresponding \zrr-matrices are
also related by (\ref{simtransrelZ}).

Lastly, the $\rho_i(x;q)$'s are determined from (\ref{rediteZ}b), which
we rewrite below
\begin{eqnarray}
\lefteqn{
\Bigl( \rho_1(x;q) \cp_1(q) + \rho_0(x;q)\cp_0(q) + \rho_N(x;q)\cp_N(q)
\Bigr)
\left( x e^V_0\otimes q^{-h_0^V/2} + q^{h_0^V/2} \otimes e_0^V\right) 
} \nonumber \\
& & = 
\left( e^V_0\otimes q^{-h_0^V/2} + q^{h_0^V/2} \otimes x e_0^V\right)
\Bigl( \rho_1(x;q) \cp_1(q) + \rho_0(x;q)\cp_0(q) + \rho_N(x;q)\cp_N(q)
\Bigr) \label{raffitZ}
\end{eqnarray}
Multiplying this equation by $\cp_1(q)$ on the left and by $\cp_N(q)$
on the right, and with the help of (\ref{propprojZ}) and (\ref{proppNZ}),
gives
\begin{eqnarray}
\lefteqn{
\rho_1(x;q) \cp_1(q)
\left( x e^V_0\otimes q^{-h_0^V/2} + q^{h_0^V/2} \otimes e_0^V\right)
\cp_N(q)
} \nonumber \\
& & \qquad\qquad =
\cp_1(q)
\left( x e^V_0\otimes q^{-h_0^V/2} + q^{h_0^V/2} \otimes e_0^V\right)
\rho_0(x;q) \cp_N(q).  \label{raffit1Z}
\end{eqnarray}
Evaluating (\ref{raffit1Z}) using the explicit expressions for
the states (\ref{qsttyZ}) and (\ref{qstatyZ}),
and the generators (\ref{afchZ}b) and (\ref{afdefchZ}), leads to the result
\zbe
\rho_1(x;q) =  \left( \frac{x - q^2}{1-x q^2} \right)\rho_0(x;q).
\label{rh1Z}
\zee
To solve for $\rho_N(x;q)$ we multiply (\ref{raffitZ}) on the
left by $\cp_1(q)$
and on the right by $\cp_0(q)$ to get
\begin{eqnarray}
\lefteqn{
\rho_1(x;q) \cp_1(q)
\left( x e^V_0\otimes q^{-h_0^V/2} + q^{h_0^V/2} \otimes e_0^V\right)
\cp_0(q)
} \nonumber \\
& & \quad\qquad =
\cp_1(q)
\left( x e^V_0\otimes q^{-h_0^V/2} + q^{h_0^V/2} \otimes e_0^V\right)
\left( \rho_0(x;q)\cp_0(q) +\rho_0(x;q) \cp_N(q)\right),
\label{raffitNZ}
\end{eqnarray}
from which it follows that
\zbe
\rho_N(x;q) = 2 q^2 \frac{q^2 - 1}{\sqrt{1+ q^4}} 
\frac{1-x^2}{(x-q^2)(1-x q^2)} \rho_0(x;q). \label{rhNZ}
\zee
Note that in calculating $\rho_{1,N}(x;q)$, the normalization of 
$e^V_0$ is irrelevant, and only the fact that $e_0^V \propto E_{41}$
is needed.
The complete \zrr-matrix satisfying (\ref{rediteZ}) is therefore (setting
$\rho_0(x;q)=1$)
\zbe
\rt(x;q) = \cp_0(q) + \left( \frac{x - q^2}{1-x q^2} \right)\cp_1(q)
+ 2 q^2 \frac{q^2 - 1}{\sqrt{1+ q^4}}\frac{1-x^2}{(x-q^2)(1-x q^2)}
\cp_N(q), \label{ybrmatZ}
\zee
and is unique up to multiplication by a scalar factor.  One can easily
show that
\zbe
\rt(1;q) = 1,\qquad \rt(x;q) \rt(x^{-1};q) = 1, \label{runZ}
\zee
where the latter relation is the unitarity requirement.  As stated above,
the \zrr-matrix for \uospa\ has been previously computed 
in \cite{massaR,goliyztsR} and \cite{deguchiR}.  
Our solution agrees with that of \cite{massaR}, where 
the same \uosp\ simple root system is used.
To obtain (\ref{ybrmatZ}) from the more general 
result presented in \cite{massaR} (in the limit $b\rightarrow 0$), one
needs to take $x\rightarrow x^{-1}$ and $q\rightarrow q^{-1}$ due to the
different conventions used.

\medskip
\noindent
{\it 4d. Crossing symmetry}

\smallskip
To go from the \zrr-matrix (\ref{ybrmatZ}) to a physical \zss-matrix we need
to calculate the overall scalar factor $v(x;q)$.  This factor is necessary in
order to make $S(x;q)$ crossing symmetric.  

As it stands, (\ref{ybrmatZ}) satisfies the Yang-Baxter
and unitarity requirements.  We have seen that these two conditions follow
from the intertwining property of $\rt(x;q)$ with the comultiplication 
$\Delta$.  In a similar manner, crossing symmetry arises due to an
additional ``intertwining'' property of the universal \zrr-matrix $R(x;q)$
with the antipode operation $\cs$.  With this additional property the
\zrr-matrix also becomes crossing symmetric, which allows us to
identify $\rt(x;q)$ with the \zss-matrix.
In terms of the notation (\ref{sformZ}), this means absorbing the
scalar factor into $\rt(x;q)$, $v(x;q) \rt(x;q)\rightarrow \rt(x;q)$,
setting $S(x;q) = \rt(x;q) = P R(x;q)$, and requiring this new \zrr-matrix
to satisfy the crossing constraint (\ref{crsrZ}) along with (\ref{rediteZ}).
In the general discussion below, we will make this change taking 
$S(x;q) = \rt(x;q)$.  We begin by reviewing the antipode operation.
{}From its relation to the
universal \zrr-matrix we will then derive the crossing symmetry constraint. 
(For further details the reader is referred to \cite{deliusR}.)

The antipode is one of the defining structures of a Hopf algebra which 
acts to connect the multiplication and comultiplication operations.
For a non-affine quantum supergroup $U_q(g)$, the antipode $\cs$ is
defined by its action on the generators as follows 
\cite{zhgobrybR} $(i=1,2,\ldots,r)$
\[
\pi_V\left( \cs(e_i)\right) = - q^{-\al_i\cdot\al_i/2} \pi_V(e_i)
\]
\[
\pi_V\left( \cs(f_i)\right) = - q^{+\al_i\cdot\al_i/2} \pi_V(f_i)
\]
\zbe
\pi_V\left( \cs(h_i)\right) = - \pi_V(h_i),
\label{naqatpZ}
\zee
where the notation $\pi_V(\,\cdot\,)$ means all quantities are taken in the
representation $V$. (In the notation of section 3d, $\pi_V(e_i) = e_i^V$,
etc..)  For $q=1$ this reduces to the classical antipode $\cs_{\rm cl}$
\zbe
\pi_V\left( \cs_{\rm cl}(a)\right) = - \pi_V(a),\ \ a\in\{e_i,f_i,h_i\}.
\label{clatpZ}
\zee
The antipode can be extended to a graded anti-automorphism, so that for
homogeneous elements $a$ and $b$ (i.e., elements with definite parity)
\zbe
\cs(ab) = (-1)^{d(a)d(b)} \cs(b)\cs(a).
\zee

To build the crossing relation a concept of charge conjugation is
required, whereby a particle is transformed into its antiparticle.  At the
more formal level, the charge conjugation operation is expressed in terms 
of a charge conjugation matrix.  If there exists a matrix $C$ satisfying
\zbe
\pi_V\left(\cs(a)\right) = C^{-1} \left( \pi_V(a)\right)^{st} C,
\ \  a\in\{e_i,f_i,h_i\}, \label{qccmZ}
\zee
then $C$ is defined to be the charge conjugation matrix.  In the classical
case this becomes
\zbe
- \pi_V(a) = C^{-1} \left( \pi_V(a)\right)^{st} C,
\ \  a\in\{e_i,f_i,h_i\}. \label{clccmZ}
\zee
Here $st$ denotes the supertranspose, which for homogeneous elements is
given by
\zbe
 \left(\pi_V(a) \right)^{st}_{\al\bt} =
(-1)^{d(\al)d(\bt)+d(\bt)} \left( \pi_V(a)\right)_{\bt\al}.
\label{sutrZ}
\zee
For the \uosp\ simple root system (\ref{srootsZ}), we have 
$\al_1\cdot\al_1 = \al_2\cdot\al_2 = 0$, and one can show that in the 
fundamental representation (\ref{chrepZ})
\zbe
C =
\left( \begin{array}{cccc}
 & & & \pm 1 \\
 & & 1 & \\
 & 1 & & \\
\mp 1 & & &
\end{array} \right).
\label{ccmZ}
\zee
The signs refer to the two different parity assignments (see section 4c).

For an affine quantum supergroup $U_q(\hat{g})$, the antipode is still
given by (\ref{naqatpZ}).  However, in order to correctly
define a charge conjugation
matrix it is necessary to shift the spectral parameter.  Thus it is
convenient to explicitly display the $\zt$ dependence $(i=0,1,\ldots,r)$
\[
\pi_V^\zt\left( \cs(e_i)\right) = - q^{-\al_i\cdot\al_i/2} 
e^{+\zt s_i} \pi_V(e_i)
\]
\[
\pi_V^\zt\left( \cs(f_i)\right) = - q^{+\al_i\cdot\al_i/2} 
e^{-\zt s_i} \pi_V(f_i)
\]
\zbe
\pi_V^\zt\left( \cs(h_i)\right) = - \pi_V(h_i).
\label{aqatpZ}
\zee
Here the notation $\pi_V^\zt(\,\cdot\,)$ indicates the $\zt$ dependence 
and $s_i$ is the Lorentz spin of $e_i$, i.e.,
\zbe
\pi_V^\zt(e_i) = x_i \pi_V(e_i) = x_i e_i^V, \qquad
\pi_V^\zt(f_i) = x_i^{-1} \pi_V(f_i) = x_i^{-1} f_i^V,\qquad
x_i = e^{\zt s_i}, \label{larepZ}
\zee
gives the affinized representation of $U_q(\hat{g})$ on a loop algebra
in some gradation $(s_0,s_1,\dots,s_r)$.  The $e_i^V$, $f_i^V$ and $h_i^V$
are representation matrices, as in (\ref{chrepZ}), (\ref{afchZ}b) and
(\ref{afdefchZ}) for \uospa.
In the homogeneous gradation we are working with $s_0 = -2/\gah, s_1=s_2=0 $,
giving
\zbe
\pi_V^\zt(e_{1,2}) = e_{1,2}^V,\qquad
\pi_V^\zt(e_0) = x e_0^V \quad (x = e^{-2\zt/\gah}),
\label{larephgZ}
\zee
and so forth.

If (\ref{aqatpZ}) can now be written as
\[
\pi_V^\zt\left(\cs(e_i)\right) = e^{+(i\pi +\zt)s_i} C^{-1}
\left(\pi_V(e_i)\right)^{st} C = 
C^{-1} \left( \pi_V^{\zt+i\pi}(e_i)\right)^{st} C
\]
\[
\pi_V^\zt\left(\cs(f_i)\right) = e^{-(i\pi +\zt)s_i} C^{-1}
\left(\pi_V(f_i)\right)^{st} C = 
C^{-1} \left( \pi_V^{\zt+i\pi}(f_i)\right)^{st} C
\]
\zbe
\pi_V^\zt\left(\cs(h_i)\right) = C^{-1} \left(\pi_V(e_i)\right)^{st} C
= C^{-1} \left( \pi_V^{\zt+i\pi}(h_i)\right)^{st} C, 
\label{qaccmZ}
\zee
for some matrix $C$, then this defines $C$ to be the (affine) charge
conjugation matrix.  Specializing to the algebra \uospa, we have
\zbe 
q = -e^{-i\pi/\gah}
\zee
\zbe
q^{\mp\al_1\cdot\al_1/2} = q^{\mp\al_2\cdot\al_2/2} = 1,\qquad
q^{\mp\al_0\cdot\al_0/2} = e^{\mp 2 i\pi/\gah},
\zee
giving for (\ref{aqatpZ})
\[
\pi_V^\zt\left(\cs(e_{1,2})\right) = - \pi_V(e_{1,2})
\]
\[
\pi_V^\zt\left(\cs(e_{0})\right) = - e^{+(\zt+i\pi)(-2/\gah)} \pi_V(e_{0}) 
\]
\[
\pi_V^\zt\left(\cs(f_{1,2})\right) = - \pi_V(f_{1,2})
\]
\[
\pi_V^\zt\left(\cs(f_{0})\right) = - e^{-(\zt+i\pi)(-2/\gah)} \pi_V(f_{0}) 
\]
\zbe
\pi_V^\zt\left(\cs(h_{0,1,2})\right) = - \pi_V(h_{0,1,2}).
\label{qospatpZ}
\zee
Comparing (\ref{qospatpZ}) with (\ref{qaccmZ}), we see that $C$
has to satisfy (\ref{clccmZ}) for all the
affine generators.  One can check the same \uosp\ conjugation
matrix (\ref{ccmZ}) also satisfies (\ref{clccmZ}) for the fundamental affine
representation (\ref{chrepZ}), (\ref{afchZ}b) and (\ref{afdefchZ}).
(The fact that the charge conjugation matrix is the same for \uosp\ and
\uospa\ is a consequence of the specific representation, gradation and
root system we are working with.  In general this is not the case.)

The crossing relation can now be derived from the following relation
between the universal \zrr-matrix and the antipode
\zbe
(\cs\otimes 1)R = R^{-1}.
\zee
Evaluating this in the representation $\pi_V^{\zt_1}\otimes\pi_V^{\zt_2}$,
and making use of (\ref{qaccmZ}) one gets $(\zt=\zt_1-\zt_2)$
\zbe
R(\zt;\gah)(C^{-1}\otimes 1)(R(i\pi + \zt;\gah))^{{st}_1} (C\otimes 1) = 1,
\label{crsrZ}
\zee
where ${st}_1$ means taking the supertranspose in only the first space of the
tensor product $V\otimes V$.  In components we have
\zbe
\left(R^{st_1}\right)_{a_1 a_2}^{b_1 b_2} = 
(-1)^{d(a_1)d(b_1) + d(a_1)} R_{b_1 a_2}^{a_1 b_2}.
\zee
We also have written $(\zt;\gah)$ instead of $(x;q)$ for the variable
dependence of $R$ and will freely use both notations.  
Equation (\ref{crsrZ}) is easily derived using the general expression
for $R$ obtained via the quantum double construction \cite{drinqdR,beleR}
\zbe
R= \sum_i a_i\otimes a^i.
\zee
Substituting $S(\zt;\gah) = PR(\zt;\gah)$ into (\ref{crsrZ})
gives the crossing relation for the \zss-matrix
\zbe
P S(\zt;\gah)(C^{-1}\otimes 1)(P S(i\pi + \zt;\gah))^{st_1} (C\otimes 1) = 1.
\label{crss1Z}
\zee
With the unitarity condition
\zbe
S(\zt;\gah) S(-\zt;\gah) = 1, \label{sunZ}
\zee
equation (\ref{crss1Z}) can be rewritten as
\zbe
S(\zt;\gah) = (C^{-1}\otimes 1)(P S(i\pi - \zt;\gah))^{st_1}(C\otimes 1)P.
\label{crss2Z}
\zee
We have derived (\ref{crss2Z}) rather formally, though its physical
interpretation is simple: the amplitude for the direct-channel process
\zbe
\zsi{a_1,\zt_1}\otimes \zsi{a_2,\zt_2} \longrightarrow
\zsi{b_2,\zt_2}\otimes \zsi{b_1,\zt_1},
\zee
is the same as the amplitude for the cross-channel process
\zbe
\zsi{a_2,\zt_2 +i\pi/2}\otimes \zsi{\overline{b}_1,\zt_1-i\pi/2} 
\longrightarrow
\zsi{\overline{a}_1,\zt_1-i\pi/2}\otimes \zsi{b_2,\zt_2+i \pi/2},
\zee
where the overbar denotes the conjugated state as determined by the
charge conjugation matrix.

\medskip
\noindent
{\it 4e. The minimal \zss-matrix}

\smallskip
We now want to build a crossing symmetric \zss-matrix starting from the
previously obtained result
\zbe
S(x;q) = v(x;q) \rt(x;q) = v(x;q) P R(x;q).
\zee
Here we have returned to to the notation of section 4c, with
$\rt(x;q)$ being the specific solution (\ref{ybrmatZ})
and not the crossing symmetric \zrr-matrix satisfying (\ref{crsrZ}).
In general, a solution for $v(x;q)$ making
$S(x;q)$ crossing symmetric will spoil unitarity.  (Recall $\rt(x;q)$ is
unitary (\ref{runZ}).)  Thus the crossing and unitarity equations need to be 
considered together when determining $v(x;q)$.

The constraints (\ref{sunZ}) and (\ref{crss1Z}) give the following equations
for $v(x;q)$
\zbe
v(x;q)v(x^{-1};q) = 1  \label{vun1Z}
\zee
\zbe
v(x;q) v(xq^2;q)\left[ R(x;q)(C^{-1}\otimes 1)(R(xq^2;q))^{{st}_1}
(C\otimes 1)\right] = 1,  \label{vcrs1Z}
\zee
where we have used
\zbe
x(\zt) = e^{-2\zt/\gah},\qquad x(i\pi+\zt) = x(\zt) q^2.
\label{xthetaZ}
\zee
Evaluating the quantity in square brackets, $[\ldots]$, we get
\zbe
[\ldots] = \frac{(x - q^2)}{(x-1)}\frac{(1-x q^2)}{(1-x q^4)},
\zee
for both parity assignments.  Therefore the functional relations
determining the scalar factor are
\zbe
v(x;q) v(x^{-1};q) = 1
\label{vuncrsZ}
\zeqnalt
\zee
\zadeqn%
\zbe
v(x;q) v(x q^2;q) = \frac{(x-1)}{(x-q^2)}\frac{(1-x q^4)}{(1- x q^2)}.
\zeqnalt
\zee
\zsetc%
\zeqnorg%
We choose to express these equations in a slightly different form.
Introduce $\vh(x;q)$ via
\zbe
v(x;q) = \frac{1}{4 \pi^2} \frac{1}{x q^2}\, (x-q^2)(1-x q^2) \vh(x;q).
\label{vvhrelZ}
\zee
Then (\ref{vuncrsZ}a) and (\ref{vuncrsZ}b) are equivalent to
\zbe
\vh(x;q)\vh(x^{-1};q) = \frac{(4\pi^2 q^2)^2}{(1-x q^2)^2 (1- x^{-1} q^2)^2}
\label{vhuncrsaZ}
\zee
\zadeqn%
\zbe
\vh(x^{-1} q^2;q) = \vh(x;q),
\zeqnalt
\zee
\zsetc%
\zeqnorg%
or in terms of $(\zt,\gah)$
\zbe
\vh(\zt;\gah)\vh(-\zt;\gah)  = 
\frac{\pi^4}{\sin^2\lb \zl(\pi + i\zt)\rb \sin^2\lb \zl(\pi-i\zt)\rb}
\label{vhuncrsbZ}
\zeqnalt
\zee
\zadeqn%
\zbe
\vh(i\pi-\zt;\gah) = \vh(\zt;\gah).
\zeqnalt
\zee
\zsetc%
\zeqnorg%

A solution to these equations can be constructed iteratively as discussed
in \cite{penatiR,leclairR}.  One begins with a 
specific solution of (\ref{vhuncrsbZ}a), call
it $\vh_0$.  This will in general not be a solution of (\ref{vhuncrsbZ}b).
One then looks for a solution of (\ref{vhuncrsbZ}b), denoted $\vh_1$, 
that is of the form $\vh_1 = \vh_0 f_1$ for some function $f_1$.
Now $\vh_1$ will no longer satisfy (\ref{vhuncrsbZ}a) 
and one has to re-construct
a unitary solution.  In order not to end up with the prior solution $\vh_0$,
a new solution $\vh_2$, of the form $\vh_2 = \vh_1 f_2 = \vh_0 f_1 f_2$, is
sought.  The process is then repeated, eventually giving a solution
in the form of an infinite product $\vh = \vh_0 \prod_i f_i$ (assuming
everything converges).  This recursive method will become clear 
as we solve (\ref{vhuncrsbZ}).  The final solution for $v$ will depend on the
choice of $\vh_0$.  This reflects the fact that (\ref{vhuncrsbZ}) does not 
have a unique solution.   However all solutions will differ only by a product
(possibly infinite) of CDD factors.  For a certain choice
of $\vh_0$ the solution
will be minimal, meaning that it will contain a minimum number of poles in
the physical strip $0< {\rm Im}\,\zt < \pi$.  We will build this
minimal solution.

In constructing a minimal solution it is more convenient, and perhaps
even necessary, to work with (\ref{vhuncrsbZ}) rewritten
in terms of gamma functions as
\zbe
\vh(\zt;\gah)\vh(-\zt;\gah)  = 
\lb \zG\lb 1 - \zl(1 + \ztf)\rb \zG\lb\zl(1+\ztf)\rb
\zG\lb 1 - \zl(1 - \ztf)\rb \zG\lb\zl(1-\ztf)\rb \rb^2
\label{vgmZ}
\zeqnalt
\zee
\zadeqn%
\zbe
\vh(i\pi-\zt;\gah) = \vh(\zt;\gah).
\zeqnalt
\zee
\zsetc%
\zeqnorg%
For an initial solution of (\ref{vgmZ}a) we  take
\zbe
\vh_0(\zt;\gah) = - \lb \zG\lb 1 - \zl(1 + \ztf)\rb
\zG\lb\zl(1-\ztf)\rb \rb^2. \label{vinZ}
\zee
However, this does not solve (\ref{vgmZ}b).  So we 
adjust (\ref{vinZ}) as follows
\zbe
\vh_1(\zt;\gah) = \vh_0(\zt;\gah) f_1(\zt;\gah),
\zee
where
\zbe
f_1(\zt;\gah) = \vh_0(i\pi-\zt;\gah).
\zee
Now $\vh_1(\zt;\gah)$ is crossing symmetric but spoils the unitarity
constraint.  Unitarity is restored by taking
\zbe
\vh_2(\zt;\gah) = \vh_0(\zt;\gah) f_1(\zt;\gah) f_2(\zt;\gah)
\zee
where
\zbe
f_2(\zt;\gah) = \frac{1}{f_1(-\zt;\gah)}.
\zee
At the next step we have, restoring crossing symmetry,
\zbe
\vh_3(\zt;\gah) = \vh_0(\zt;\gah) f_1(\zt;\gah) f_2(\zt;\gah) f_3(\zt;\gah)
\zee
where
\zbe
f_3(\zt;\gah) = f_2(i\pi-\zt;\gah),
\zee
and so forth, with the process never terminating.  
{}From the structure of the $\vh_i$'s, we see that the complete solution
to (\ref{vgmZ}) takes the form
\zbe
\vh(\zt;\gah)  = \vh_0(\zt;\gah) \prod_{n=1}^\infty
\frac{\vh_0((2n-1)\, i\pi - \zt;\gah)}{\vh_0((2n-1)\, i \pi + \zt;\gah)}
\frac{\vh_0(2n\, i\pi + \zt;\gah)}{\vh_0(2n\, i \pi - \zt;\gah)}.
\zee
Explicitly we have
\zbe
\vh(\zt;\gah) = -\zG^2\lb\zl(1 + \ztf)\rb \zG^2\lb 1 - \zl(1+\ztf)\rb
\lb \frac{\zG\lb 1 + \zl \ztf \rb}{\zG\lb 1 - \zl \ztf \rb} \rb^2
\, \lb I(\zt;\gah)\rb^2, \label{vhsolZ}
\zee
where
\zbe
I(\zt;\gah) = \prod_{n=1}^\infty
\frac{\zG\lb 2n \zl + \zl\ztf\rb}{\zG\lb 2n \zl - \zl\ztf\rb}
\frac{\zG\lb 1+ 2n \zl + \zl\ztf\rb}{\zG\lb 1+2n \zl - \zl\ztf\rb}
\frac{\zG\lb (2n-1) \zl - \zl\ztf\rb}{\zG\lb (2n-1) \zl + \zl\ztf\rb}
\frac{\zG\lb 1+ (2n-1) \zl - \zl\ztf\rb}{\zG\lb 1+ (2n-1) \zl + \zl\ztf\rb}.
\label{IfZ}
\zee
The factor $I(\zt;\gah)$ is identical to  
the infinite product that appears 
in the sine-Gordon \zss-matrix \cite{zzsgR}.  In the notation of 
\cite{zzsgR} (identifying
$1/\gah = 8\pi/\gamma$)
\zbe
I(\zt;\gah) = \prod_{n=1}^\infty R_n(\zt) R_n(i\pi-\zt)  
\zee
This also shows that (\ref{IfZ}) converges.
Taking into account (\ref{vvhrelZ}), the final
expression for $v(\zt;\gah)$ that solves (\ref{vuncrsZ}) is
\zbe
v(\zt;\gah) = \frac{\zG\lb 1 - \zl - \zl\ztf\rb}{\zG\lb 1 - \zl + \zl\ztf\rb}
\frac{\zG\lb \zl + \zl\ztf\rb}{\zG\lb \zl - \zl\ztf\rb}
\lb \frac{\zG\lb 1 + \zl\ztf\rb}{\zG\lb 1-\zl\ztf\rb} \rb^2
\, \lb I(\zt;\gah)\rb^2. \label{vsolZ}
\zee
Therefore, the minimal \zss-matrix for the Toda model (\ref{dtacZ}) is 
\zbe
S(\zt;\gah) = v(\zt;\gah) \rt(\zt;\gah), \label{minSZ}
\zee
with $\rt(\zt;\gah)$ given by (\ref{ybrmatZ}) and $v(\zt;\gah)$ as above.
The \zss-matrix is unique up to an arbitrariness only of the CDD type.
As previously stated, the CDD factors are determined by applying the bootstrap
procedure.  In general this is a complicated task and requires knowing all the
particle multiplets of the theory.  The bootstrap equation constrains the
poles and zeros of the various \zss-matrices associated with the different 
multiplets.  The final result is that the complete \zss-matrix (for the vector
representation) takes the form
\zbe
S_{\rm complete}(\zt;\gah) = X(\zt;\gah) v(\zt;\gah) \rt(\zt;\gah),
\zee
where $X(\zt;\gah)$ is a product of CDD type factors.

We give alternative expressions for the $\rho_i(\zt;\gah)$ factors
(\ref{rh1Z}) and (\ref{rhNZ}) in terms of gamma functions
\zbe
\rho_1(\zt;\gah) = 
\frac{\zG\lb 1 - \zl +\zl\ztf\rb \zG \lb\zl - \zl\ztf\rb}
{\zG\lb 1 - \zl -\zl\ztf\rb \zG\lb\zl+\zl\ztf\rb}
\label{rh1gmZ}
\zee
\zbe
\rho_N(\zt;\gah) = - \frac{2 q}{\sqrt{1+ q^4}}
\frac{\zG\lb 1 - \zl -\zl\ztf\rb \zG\lb \zl + \zl\ztf\rb
\zG\lb 1 - \zl +\zl\ztf\rb \zG\lb\zl -\zl\ztf\rb}{\zG\lb 1 - \zl\rb
\zG\lb\zl\rb \zG\lb 1 - \frac{2}{\gah}\ztf\rb \zG\lb \frac{2}{\gah}\ztf\rb}.
\label{rhNgmZ}
\zee
These expressions are useful for analyzing the pole structure of
$S(\zt;\gah)$ (see section 5).

\medskip
\noindent
{\it 4f.  The Yangian limit and the \osp\ current-current \zss-matrix}

\smallskip
In this section we compute the Yangian symmetric \zss-matrix, 
$S_Y(\zt)$, for the original \osp\ current-current model (\ref{seffVZ}).
As explained in section 2, $S_Y(\zt)$ can be obtained from (\ref{minSZ})
by taking the marginal limit $\bh\rightarrow 1^+$, or $\ep\rightarrow 0^+$
where $\ep\equiv 1/\gah$.  (This limit is equivalent to the rational
limit usually considered for \zrr-matrices, and the
resulting \zrr-matrix being referred to as the rational \zrr-matrix.)

The marginal limit is easily evaluated for the various components of
$S(\zt;\gah)$.  For the projectors, which are well defined as
$\ep\rightarrow 0^+$, we simply set $q=-1$.  For the remaining factors,
$v(\zt;\gah)$ and $\rho_{1,N}(\zt;\gah)$, we make use of
\zbe
\lim_{\ep\rightarrow 0^+} \lb \frac{\zG(\ep A)}{\zG(\ep B)} \rb
= \frac{B}{A}.
\zee
One finds 
\begin{eqnarray}
\lim_{\ep\rightarrow 0^+} v(\zt;\gah) & = &
\frac{1 -\ztf}{1 + \ztf} \lb \prod_{n=1}^\infty 
\frac{\lb 2n-1 + \ztf\rb }{\lb 2n-1-\ztf\rb} 
\frac{\lb 2n-\ztf\rb}{\lb 2n+\ztf\rb} \rb^2 \nonumber \\
& = & \frac{1 -\ztf}{1 + \ztf}
\lb \frac{\zG\lb 1 +\fa\ztf\rb \zG\lb \fa - \fa\ztf\rb}{\zG\lb 1 -\fa\ztf\rb
\zG\lb\fa + \fa\ztf\rb} \rb^2
\end{eqnarray}
\zbe
\lim_{\ep\rightarrow 0^+} \rho_1(\zt;\gah) = \frac{1+\ztf}{1-\ztf}
\zee
\zbe
\lim_{\ep\rightarrow 0^+} \rho_N(\zt;\gah) = 2\sqrt{2}
\frac{\ztf}{(1+\ztf)(1-\ztf)}.
\zee
These results could have been obtained directly from (\ref{rh1Z})
and (\ref{rhNZ}), and
by taking the marginal limit before calculating $v(\zt)$.
Combining everything, $S_Y(\zt)$ takes the block diagonal form as in
(\ref{bldfZ}) with the individual blocks being:\newline
(i) One-dimensional blocks:
\zbe
S_Y(\zt) = I_Y^2(\zt),
\label{minSYZ}
\zeqnalt
\zee
for  $\ztp{1}{1}$ and $\ztp{4}{4}$; and
\zadeqn%
\zbe
S_Y(\zt) = \frac{i\pi+\zt}{i\pi-\zt}\, I_Y^2(\zt),
\zeqnalt
\zee
for $\ztp{2}{2}$ and $\ztp{3}{3}$. \newline
(ii) Two-dimensional blocks:
\zadeqn%
\zbe
S_Y(\zt) = \frac{1}{i\pi-\zt} \left( 
\begin{array}{cc}
i\pi & \pm \zt \\
\pm\zt & i\pi 
\end{array}
\right)\, I_Y^2(\zt),
\zeqnalt
\zee
for the four pairs of states 
\[
(\ztp{1}{2},\ztp{2}{1}), \ \
(\ztp{1}{3},\ztp{3}{1}), \ \ 
(\ztp{2}{4},\ztp{4}{2}), \ \ 
(\ztp{3}{4},\ztp{4}{3}).
\]
(ii) Four-dimensional block:
\zadeqn%
\zbe
S_Y(\zt) = -\frac{1}{(i\pi-\zt)^2} \left(
\begin{array}{cccc}
\pi(\pi+2i\zt)  & \pm i\pi\zt & \pm i\pi\zt & - \zt^2 \\
\mp i\pi\zt & \pi^2 & \zt(\zt-2 i\pi) & \pm i\pi\zt \\
\mp i\pi\zt & \zt(\zt- 2i\zt) & \pi^2 & \pm i\pi\zt \\
-\zt^2 & \mp i\pi\zt & \mp i\pi\zt & \pi(\pi+ 2 i\zt) 
\end{array}
\right) \, I_Y^2(\zt),
\zeqnalt
\zee
\zsetc%
\zeqnorg%
for the basis states
\[
(\ztp{1}{4},\ztp{2}{3},\ztp{3}{2},\ztp{4}{1}).
\]
The factor $I_Y(\zt)$ is the $\ep\rightarrow 0^+$ limit of $I(\zt;\gah)$: 
\zbe
I_Y(\zt) = \frac{\zG\lb 1 + \frac{i\zt}{2\pi} \rb
\zG\lb\fa -\frac{i\zt}{2\pi} \rb}{\zG\lb 1 - \frac{i\zt}{2\pi} \rb
\zG\lb\fa +\frac{i\zt}{2\pi} \rb}. \label{IYfZ}
\zee
This completes our calculation of the \zss-matrix for the \osp\ 
current-current model (\ref{seffVZ}).

We have not yet proven that (\ref{minSYZ}) is
Yangian symmetric.  This means showing
$S_Y(\zt)$ commutes with the comultiplication of the Yangian generators.
Unfortunately the general theory of super Yangians associated with Lie
superalgebras is not as fully developed as that of quantum supergroups.
(See \cite{nazarovR,zhangsyR} and 
references therein for a discussion of super Yangians.)
For ordinary Lie algebras, one can realize a Yangian structure in terms of
non-local charges arising from curvature-free currents \cite{berhyR}.  
This construction
is particularly useful is studying Yangian symmetric field theories and
\zss-matrices.  However we are unaware of a similar characterization of
the Yangians for Lie superalgebras.  Thus we do not
strictly show (\ref{minSYZ}) to
be Yangian symmetric but give some supporting evidence.  Our discussion
is based on a comparison with bosonic Yangian symmetric systems.  Examples
of such systems include Gross-Neveu type models and the sine-Gordon model.
At the marginal point, these models consist of current-current type
perturbations of a free field theory.  Furthermore, their Yangian symmetric
\zss-matrices can be obtained by taking the marginal limit of a \zss-matrix
with affine quantum group symmetry \cite{leclairR}.  
(For the Gross-Neveu model this quantum group symmetric
\zss-matrix corresponds to an affine Toda theory.)  The Yangian
symmetry of the \zss-matrix, i.e., the vanishing commutator of $S$ with the
Yangian comultiplication, follows from the marginal limit of the
quantum affine symmetry relations.  Here we are dealing with a supersymmetric
analog of the bosonic Gross-Neveu model.  The
current-current model (\ref{seffVZ}) is
a Gross-Neveu type model based on \osp, and (\ref{todaacZ}) is
an \ospa\ Toda-type system.  As for the bosonic case, we expect
that the marginal \zss-matrix is Yangian symmetric and the
corresponding symmetry relations can be extracted from (\ref{scomrelZ}).

Let us first recall the bosonic situation.  The Yangian $Y(g)$ based on a
semi-simple Lie algebra $g$ of rank $r$ is generated by the set of charges
$\{ Q_a^{(0)},Q_a^{(1)}\}_{a=1,\ldots, r}$ satisfying, among others, the
relations \cite{berhyR,beleR,drinfeldR,beriyR}
\[
[ Q^{(0)}_a, Q^{(0)}_b] = f_{abc} Q^{(0)}_c
\]
\zbe
[Q^{(0)}_a,Q^{(1)}_b] = f_{abc} Q^{(0)}_c, \label{YalgZ}
\zee
where the $f_{abc}$'s are structure constants of $g$.  $Y(g)$ can be
made into a Hopf algebra with the comultiplication
\[
\zD(Q^{(0)}_a) = Q^{(0)}_a\otimes 1 + 1\otimes Q^{(0)}_a
\]
\zbe
\zD(Q^{(1)}_a) = Q^{(1)}_a\otimes 1 + 1\otimes Q^{(1)}_a
- \fa f_{abc} Q^{(0)}_b\otimes Q^{(0)}_c.
\label{YahaZ}
\zee
The charges have non-trivial Lorentz spin and under a Lorentz boost by $\zt$,
denoted $T_\zt$, behave as
\[
T_\zt(Q^{(0)}_a) = Q^{(0)}_a
\]
\zbe
T_\zt(Q^{(1)}_a) = Q^{(1)}_a + c\zt Q^{(0)}_a,
\label{YalsZ}
\zee
where $c$ is a normalization constant independent of the index $a$.  The
Yangian symmetric \zss-matrix then commutes
with (\ref{YahaZ}) evaluated in the
``gradation'' given by $T_\zt$
\zbe
S(\zt_1-\zt_2) (T_{\zt_1}\otimes T_{\zt_2})\zD(Q^{(0,1)}_a) =
(T_{\zt_2}\otimes T_{\zt_1})\zD(Q^{(0,1)}_a) S(\zt_1-\zt_2),
\zee
or explicitly
\zbe
S(\zt_1-\zt_2) \lb Q^{(0)}_a\otimes 1 + 1\otimes Q^{(0)}_a \rb =
\lb Q^{(0)}_a\otimes 1 + 1\otimes Q^{(0)}_a \rb S(\zt_1-\zt_2),
\label{YasysZ}
\zeqnalt
\zee
\zadeqn%
\begin{eqnarray}
\lefteqn{ S(\zt_1-\zt_2)\lb (Q^{(1)}_a + c \zt_1 Q^{(0)}_a) 
\otimes 1 + 1\otimes (Q^{(1)}_a + c \zt_2 Q^{(0)}_a) 
- \fa f_{abc} Q^{(0)}_b\otimes Q^{(0)}_c \rb } \nonumber \\ 
& & = \lb (Q^{(1)}_a + c \zt_2 Q^{(0)}_a) \otimes 1 + 1\otimes 
(Q^{(1)}_a + c \zt_1 Q^{(0)}_a) 
- \fa f_{abc} Q^{(0)}_b\otimes Q^{(0)}_c \rb S(\zt_1-\zt_2).
\zeqnalt
\end{eqnarray}
\zsetc%
\zeqnorg%
For the bosonic $su(N)$ Gross-Neveu model,
it was shown in \cite{leclairR} that (\ref{YasysZ}) can be recovered by
expanding the affine quantum group symmetry relations in $\ep$, with
$\ep\rightarrow 0^+$ being the marginal limit.  The zeroth order term gives
the constraint (\ref{YasysZ}a) and 
the first order term leads to (\ref{YasysZ}b).

Now we show that relations similar to (\ref{YasysZ}) can also be obtained from
(\ref{scomrelZ}) as $\ep\rightarrow 0^+$ ($\ep=1/\gah$).
(Note that in writing (\ref{scomrelZ}) we had
canceled the non-zero factors $c_i$ (\ref{c2Z}) from both sides.  However as
$\ep\rightarrow 0^+$, $c_i$ seems to blow up.  As explained earlier to make
sense of (\ref{scomrelZ}) we need to regularize
the charges by taking $g\rightarrow 0$
such that $g\gah$ is finite.)  We will only display the relations for the
$e_i$'s, with those for the other generators treated similarly.  The
\uospa\ generators will be denoted as $e_{iq}, h_{iq}$ and the
$\ep$-independent \ospa\ generators as $e_i, h_i$.  (We also
drop the $V$ superscript.) {}From (\ref{chrepZ}), (\ref{afchZ}b) and
(\ref{afdefchZ}) we have to lowest order in $\ep$
\zbe
e_{1,2q} = e_{1,2}, \qquad h_{iq} = h_{i} \label{YaeZ}
\zee
\zbe
e_{0q} = i e_0 - \ep^2 \frac{\pi^2}{4} i e_0. \label{Yae0Z}
\zee
Of course (\ref{YaeZ}) is exact since these generators are not deformed.
The $i$ factor in (\ref{Yae0Z}) is a result of the Yangian point being $q=-1$
instead of $q=1$.  Also the spectral parameters are 
\zbe
x_{i0} = e^{-2\zt_i \ep} \approx 1 - 2\zt_i \ep.
\zee
Substituting these expressions into (\ref{scomrelZ}a) and (\ref{scomrelZ}b)
we find to zeroth order in $\ep$
\zbe
S_Y(\zt)\lb e_i\otimes (-1)^{-h_i/2} + (-1)^{-h_i/2}\otimes e_i\rb
= \lb e_i\otimes (-1)^{-h_i/2} + (-1)^{-h_i/2}\otimes e_i\rb S_Y(\zt),
\ \ (i=0,1,2). \label{YazesZ}
\zee
To first order in $\ep$ we have
\zbe
S_Y(\zt) \zD_{\zt_1 \zt_2}(e_i) = \zD_{\zt_2 \zt_1}(e_i) S_Y(\zt),
\label{YafisZ}
\zee
where
\zbe
\zD_{\zt_1 \zt_2}(e_{1,2}) = e_{1,2}\otimes (-1)^{-h_{1,2}/2} h_{1,2} -
(-1)^{-h_{1,2}/2} h_{1,2} \otimes e_{1,2}
\label{Yacom1Z}
\zee
\zbe
\zD_{\zt_1 \zt_2}(e_{0}) = \frac{4\zt_1}{i\pi} e_0\otimes (-1)^{-h_0/2} +
(-1)^{-h_0/2}\otimes \frac{4\zt_2}{i\pi} e_0 - 
\lb e_{0}\otimes (-1)^{-h_{0}/2} h_{0} -
(-1)^{-h_{0}/2} h_{0} \otimes e_{0} \rb.
\label{Yacom2Z}
\zee
We see that these equations agree with (\ref{YasysZ}) if we identify
\zbe
e_i^{(0)} = e_i,\qquad h_i^{(0)} = h_i
\zee
\zbe
e_i^{(1)} = 0,
\zee
and define the automorphism $T_\zt$ as
\zbe
T_\zt(e_i^{(1)}) = e_i^{(1)} - \frac{2\zt}{i\pi} s_i e_i^{(0)},
\zee
where $s_i/\gah$ is the Lorentz spin of the charge $\qbz{i}$.  Even 
though the charges $e_i^{(1)}$ are identically zero in this representation,
there is non-trivial structure due to
the comultiplication (\ref{Yacom1Z}) and (\ref{Yacom2Z}).
That the $e_i^{(1)}$'s vanish is not surprising since \uospa\ is
not deformed to order $\ep$.  Unlike (\ref{YalsZ}),
the $\zt$ dependence of $T_\zt(e_i^{(1)})$ is
not the same for all generators, which reflects the choice of gradation
(homogeneous) and the root structure of \ospa.  The twisting factors
$(-1)^{\pm h_i/2}$ are again a consequence of the $q\rightarrow -1$ limit
rather than $q\rightarrow 1$.  Off shell, these factors correspond to a
choice of statistical Klein factors and are not expected to be
dynamical \cite{beleR}.  (Taking $q\rightarrow 1$ gives an equivalent 
crossing symmetric \zss-matrix, with the same charge conjugation properties
and differing from (\ref{minSYZ}) only by a similarity transformation.)
The \zss-matrix (\ref{minSYZ}) therefore satisfies symmetry relations
analogous to those of (bosonic) Yangian symmetric \zss-matrices, which
supports the claim that $S_Y(\zt)$ has super Yangian symmetry.

\addtocounter{section}{1}
\setcounter{equation}{0}

\section*{5.  The analytic structure of $S(\zt;\gah)$}

We have argued that in the marginal limit the Toda \zss-matrix $S(\zt;\gah)$ gives
the minimal \zss-matrix for the \osp\ current-current model (\ref{seffVZ})
(in the fundamental vector representation).  As $\bh\rightarrow 1$, the
Toda theory renormalizes to the current-current 
model and $S(\zt;\gah)$ reduces to $S_Y(\zt)$.  We now study the pole structure of
$S(\zt;\gah)$.  


In the above calculation of $S(\zt;\gah)$ we have taken $\bh$ to lie
in the range $1\leq \bh<\infty$, which restricts $1/\gah$ to
\zbe
         0 \leq \frac{1}{\gah} < 1.  \label{ghrangeZ}
\zee
For these values the \zss-matrix does not have any poles in the physical 
strip $0 < {\rm Im}\,\zt <\pi$.  In particular, the physical 
\zss-matrix $S_Y(\zt)$ does not contain any bound states.
This structure is identical to that of the sine-Gordon \zss-matrix for
the range $\sqrt{8\pi} \leq \bt < \infty$ or $-1 < 8\pi/\ga \leq 0$, with
$8\pi/\ga = 0$ being the Yangian point (though in this range the sine-Gordon
model is not well-defined \cite{zzsgR}).  Of course $S(\zt;\gah)$ satisfies
the scattering constraints for all values of $1/\gah$ and we made no 
direct use of (\ref{ghrangeZ}) in deriving $S(\zt;\gah)$.  
For the sine-Gordon model the physically
relevant parameter range is $8\pi/\ga > 0$.  Analogously, we will consider
the range $1/\gah >1$.  This necessarily means 
ignoring the relation (\ref{defgaZ})
and treating $\gah$ as an independent free parameter.  Thus we are now 
viewing $S(\zt;\gah)$ as the fundamental \zss-matrix for some theory.  
(Since the quantum group symmetry used to derive $S(\zt;\gah)$ is valid
only for the range (\ref{ghrangeZ}), we cannot consider $S(\zt;\gah)$ to
be the \zss-matrix for the Toda theory (\ref{dtacZ}) for $1/\gah > 1$.
Also if $1/\gah > 1$, then (\ref{defgaZ}) implies $\bh$ is purely
imaginary, which leads to $\zD(g) = \overline{\zD}(g) < 0$, i.e., the
perturbation becomes irrelevant.  Even so, it is
interesting to study the pole structure of $S(\zt;\gah)$ independent of
any Lagrangian formulation.)
For simplicity we will restrict our analysis to generic values of $\gah$
such that $q\neq \pm 1, \pm i$, i.e., we take $1/\gah > 1$ with $1/\gah$
not an integer or half-integer.

The $\zt$-dependence is contained in the factors $v(\zt;\gah)$, 
$\rho_1(\zt;\gah)$ and $\rho_N(\zt;\gah)$.  
{}From (\ref{IfZ}) one can check that the
infinite product $I(\zt;\gah)$ contains no zeros or poles in the physical
strip.  In the remaining finite number of gamma functions, 
there are only two with zeros or poles in the physical strip
\zbe
\zG\lb 1 + \zl \ztf \rb : \qquad {\rm has\ simple\ poles\ at\ }
\zt = i\pi\gah (m+1); m = 0,1,\ldots,< \zl-1  \label{spga1Z}
\zee
and
\zbe
\zG\lb 1 - \zl - \zl \ztf \rb : \qquad {\rm has\ simple\ poles\ at\ }
\zt = i\pi - i\pi\gah (m+1); m = 0,1,\ldots,< \zl-1.  \label{spga2Z}
\zee
Combining this with (\ref{vsolZ}), (\ref{rh1gmZ}) and (\ref{rhNgmZ}), 
we find the following pole structure
for the non-zero amplitudes (in all cases $m=0,1,\ldots,<1/\gah - 1$):

\noindent
(i) amplitudes $S_{11}^{11} = S_{44}^{44}$: \newline
(a) double poles at $\zt = i\pi\gah(m+1)$ with (omitting the indices)
\zbe
S(\zt;\gah) \sim - \frac{(\pi \gah)^2 I_0^m(\gah)}{(m!(m+1)!)^2} 
\,\frac{1}{(\zt - \zt_0^m)^2}, \label{spiZ}
\zee
where
\zbe
\zt_0^m = i\pi\gah(m+1),\qquad I_0^m(\gah) = I(\zt=\zt_0^m;\gah),
\label{thI0mZ}
\zee
(b) no poles at $\zt = i\pi - i\pi\gah (m+1)$.
\newline
(ii) amplitudes $S_{22}^{22} = S_{33}^{33}$: \newline
(a) double poles at $\zt = i\pi\gah(m+1)$ with 
\zeqnalt%
\zbe
S(\zt;\gah) \sim - \frac{(\pi \gah)^2 I_0^m(\gah)}{(m!(m+1)!)^2} 
\,\frac{1}{(\zt - \zt_0^m)^2},
\zee
(b) single poles at $\zt = i\pi - i\pi\gah(m+1)$ with
\zadeqn%
\zbe
S(\zt;\gah) \sim -i\gah\sin(2\pi/\gah) 
\lb \frac{\zG \lb 2-\zl + m \rb}{\zG \lb \zl - m \rb} \rb^2 I_1^m(\gah)
\,\frac{1}{(\zt - \zt_1^m )}, \label{spiibZ}
\zee
\zsetc%
\zeqnorg%
where
\zbe
\zt_1^m = i\pi-i\pi\gah(m+1),\qquad I_1^m(\gah) = I(\zt=\zt_1^m;\gah).
\label{thI1mZ}
\zee
Using
\zbe
I(i\pi-\zt;\gah) =  I(\zt;\gah) 
\lb \frac{ \zG\lb 1+\zl +\zl \ztf\rb \zG\lb \zl + \zl \ztf\rb}{ \zG
\lb 1 - \zl\ztf\rb \zG \lb - \zl \ztf\rb} \rb^2,
\label{IcrsrelZ}
\zee
expression (\ref{spiibZ}) can be rewritten as
\setcounter{zeqn}{3}\addtocounter{equation}{-3}%
\zeqnalt
\zbe
S(\zt;\gah) \sim -\frac{2 i \pi^2 \gah I_0^m(\gah)}{(m!(m+1)!)^2}
\cot(\pi/\gah)\, \frac{1}{(\zt-\zt_1^m)}.
\zee
\addtocounter{equation}{2}\zsetc%
(iii) amplitudes $S_{12}^{12} = S_{13}^{13} = S_{24}^{24} = S_{34}^{34}$:
\newline
(a) double poles at $\zt = i\pi\gah(m+1)$ with 
\zbe
S(\zt;\gah) \sim - \frac{(\pi \gah)^2 I_0^m(\gah)}{(m!(m+1)!)^2} 
\,\frac{1}{(\zt - \zt_0^m)^2},
\zee
(b) single poles at $\zt = i\pi - i\pi\gah(m+1)$ with
\zadeqn%
\zbe
S(\zt;\gah) \sim - \frac{i\pi^2\gah I_0^m(\gah)}{(m!(m+1)!)^2}
(\cot(\pi/\gah) - i)\, \frac{1}{(\zt - \zt_1^m)}.
\zee
\zsetc%
(iv) amplitudes $S_{21}^{21} = S_{31}^{31} = S_{42}^{42} = S_{43}^{43}$:
\newline
(a) double poles at $\zt = i\pi\gah(m+1)$ with 
\zbe
S(\zt;\gah) \sim - \frac{(\pi \gah)^2 I_0^m(\gah)}{(m!(m+1)!)^2} 
\,\frac{1}{(\zt - \zt_0^m)^2},
\zee
(b) single poles at $\zt = i\pi - i\pi\gah(m+1)$ with
\zadeqn%
\zbe
S(\zt;\gah) \sim - \frac{i\pi^2\gah I_0^m(\gah)}{(m!(m+1)!)^2}
(\cot(\pi/\gah) + i)\, \frac{1}{(\zt - \zt_1^m)}.
\zee
\zsetc%
(v) amplitudes $S_{21}^{12} = S_{12}^{21} = S_{31}^{13} = S_{13}^{31} =
S_{42}^{24} = S_{24}^{42} = S_{43}^{34} = S_{34}^{43}$: 
\newline
(a) single poles at $\zt = i\pi\gah(m+1)$ with
\zbe
S(\zt;\gah) \sim \pm \frac{i\pi^2\gah I_0^m(\gah)}{(m!(m+1)!)^2}
\csc(\pi/\gah) \, \frac{1}{(\zt - \zt_0^m)},
\zee
(b) single poles at $\zt = i\pi- i\pi\gah(m+1)$ with
\zadeqn%
\zbe
S(\zt;\gah) \sim \mp \frac{i\pi^2\gah I_0^m(\gah)}{(m!(m+1)!)^2}
\csc(\pi/\gah) \, \frac{1}{(\zt - \zt_1^m)}.
\zee
\zsetc%
(vi) amplitude $S_{14}^{14}$:
\newline
(a) double poles at $\zt= i\pi \gah (m+1)$ with
\zbe
S(\zt;\gah) \sim - \frac{(\pi \gah)^2 I_0^m(\gah)}{(m!(m+1)!)^2} 
\, \frac{1}{(\zt - \zt_0^m)^2},
\zee
(b) double poles at $\zt= i\pi -i\pi \gah (m+1)$ with
\zadeqn%
\zbe
S(\zt;\gah) \sim \frac{(\pi \gah)^2 I_0^m(\gah)}{(m!(m+1)!)^2} 
\, \frac{1}{(\zt - \zt_1^m)^2}.
\zee
\zsetc%
(vii) amplitude $S_{41}^{41}$:
\newline
(a) double poles at $\zt= i\pi \gah (m+1)$ with
\zbe
S(\zt;\gah) \sim - \frac{(\pi \gah)^2 I_0^m(\gah)}{(m!(m+1)!)^2} 
\, \frac{1}{(\zt - \zt_0^m)^2},
\zee
(b) double poles at $\zt= i\pi -i\pi \gah (m+1)$ with
\zadeqn%
\zbe
S(\zt;\gah) \sim \frac{(\pi \gah)^2 I_0^m(\gah)}{(m!(m+1)!)^2} 
\, \frac{1}{(\zt - \zt_1^m)^2}.
\zee
\zsetc%
(viii) amplitudes $S_{23}^{23} = S_{32}^{32}$:
\newline
(a) double poles at $\zt= i\pi \gah (m+1)$ with
\zbe
S(\zt;\gah) \sim - \frac{(\pi \gah)^2 I_0^m(\gah)}{(m!(m+1)!)^2} 
\, \frac{1}{(\zt - \zt_0^m)^2},
\zee
(b) double poles at $\zt= i\pi -i\pi \gah (m+1)$ with
\zadeqn%
\zbe
S(\zt;\gah) \sim -\frac{(\pi \gah)^2 I_0^m(\gah)}{(m!(m+1)!)^2} 
\, \frac{1}{(\zt - \zt_1^m)^2}.
\zee
\zsetc%
(ix) amplitudes $S_{23}^{14} = S_{32}^{14} = -S_{14}^{23} = -S_{14}^{32}$:
\newline
(a) single poles at $\zt = i\pi\gah (m+1)$ with (here $S = S_{23}^{14}$)
\zbe
S(\zt;\gah) \sim \mp \frac{i\pi^2\gah I_0^m(\gah)}{(m!(m+1)!)^2}
(\cot(\pi/\gah) + i)\, \frac{1}{(\zt - \zt_0^m)},
\zee
(b) double poles at $\zt= i\pi - i\pi\gah (m+1)$ with
\zadeqn%
\zbe
S(\zt;\gah) = \sim \pm \frac{(\pi\gah)^2 I_0^m(\gah)}{(m!(m+1)!)^2}
\, \frac{1}{(\zt - \zt_1^m)^2}.
\zee
\zsetc%
(x) amplitudes $S_{23}^{41} = S_{32}^{41} = -S_{41}^{23} = -S_{41}^{32}$:
\newline
(a) single poles at $\zt = i\pi\gah (m+1)$ with (here $S = S_{23}^{41}$)
\zbe
S(\zt;\gah) \sim \pm \frac{i\pi^2\gah I_0^m(\gah)}{(m!(m+1)!)^2}
(\cot(\pi/\gah) - i)\, \frac{1}{(\zt - \zt_0^m)},
\zee
(b) double poles at $\zt= i\pi - i\pi\gah (m+1)$ with
\zadeqn%
\zbe
S(\zt;\gah) = \sim \mp \frac{(\pi\gah)^2 I_0^m(\gah)}{(m!(m+1)!)^2}
\, \frac{1}{(\zt - \zt_1^m)^2}.
\zee
\zsetc%
(xi) amplitudes $S_{32}^{23} = S_{23}^{32}$:
\newline
(a) single poles at $\zt = i\pi\gah (m+1)$:
\zbe
S(\zt;\gah) \sim \frac{2 i \pi^2 \gah I_0^m(\gah)}{(m!(m+1)!)^2}
\cot(\pi/\gah)\, \frac{1}{(\zt-\zt_0^m)},
\zee
(b) double poles at $\zt = i\pi-i\pi\gah (m+1)$ with 
\zadeqn%
\zbe
S(\zt;\gah) \sim -\frac{(\pi\gah)^2  I_0^m(\gah)}{(m!(m+1)!)^2}
\, \frac{1}{(\zt-\zt_1^m)^2}.
\zee
\zsetc%
\zeqnorg%
(xii) amplitudes $S_{41}^{14} = S_{14}^{41}$:
\newline
(a) no poles at $\zt = i\pi \gah (m+1)$,
\newline
(b) double poles at $\zt = i\pi - i\pi \gah (m+1)$ with
\zbe
S(\zt;\gah) \sim -\frac{(\pi\gah)^2  I_0^m(\gah)}{(m!(m+1)!)^2}
\, \frac{1}{(\zt-\zt_1^m)^2}.
\zee
These poles and expansions for $S(\zt;\gah)$ agree with the crossing 
constraint, which in components takes the form
\zbe
S_{a_1 a_2}^{b_2 b_1}(\zt;\gah) = (-1)^{\delta_{a_1 4} + \delta_{b_1 4}}
S_{a_2 \bar{b}_1}^{\bar{a}_1 b_2}(i\pi - \zt;\gah),\ \ 
{\rm if\ 1,4\ are\ even\ and\ 2,3\ are\  odd}
\label{crscompZ}
\zeqnalt
\zee
\zadeqn%
\zbe
S_{a_1 a_2}^{b_2 b_1}(\zt;\gah) = (-1)^{\delta_{a_1 1} + \delta_{b_1 1}}
S_{a_2 \bar{b}_1}^{\bar{a}_1 b_2}(i\pi - \zt;\gah),\ \ 
{\rm if\ 1,4\ are\ odd\ and\ 2,3\ are\  even.}
\zeqnalt
\zee
\zsetc%
\zeqnorg%
The $(-1)^\delta$ factor is due to the negative sign in the charge conjugation
matrix and the bar denotes the conjugated state ($\bar{1} = 4$, etc.).  
Equation (\ref{crscompZ}) implies that for every pole at $\zt$ in the 
direct-channel there is a corresponding pole in the cross-channel.

The amplitudes (vi)-(xii) correspond to transitions between states with zero
topological charge, $(T^1,T^2) = (0,0)$.  Simple poles in these amplitudes
at $\zt= i\pi\gah (m+1)$ can be interpreted as charge neutral bound
states in the direct-channel.  These are the ``breathers'' of the theory.
The associated cross-channel poles occur at $\zt= i\pi- i\pi\gah (m+1)$
in (ii) - (iv).  There are also bound states of charge $(+2,0), (-2,0),
(0,+2)$ and $(0,-2)$.  These ``breathing solitons'' appear as simple poles
in both the direct- and cross-channels in (v).  Lastly, there are various
double poles which probably have an explanation in terms of a Coleman-Thun
type mechanism \cite{colemanR}.

\addtocounter{section}{1}
\setcounter{equation}{0}

\section*{6. Conclusions}

We have computed the \zss-matrix for a certain disordered system.  After
disorder averaging, the theory can be written as a current-current
perturbation of an \osp\ supersymmetric CFT.  This current-current model
is known to be Yangian symmetric.  Instead of directly constructing the
Yangian symmetric \zss-matrix, we followed the approach of \cite{leclairR}.  
This approach consisted of working with a Toda-type theory which renormalizes
to the \osp\ current-current model at the marginal point.  For the Toda
theory we built  quantum group charges satisfying the \uospa\
algebra.  The Hopf algebraic structure of \uospa\ was 
then used to construct the exact \zss-matrix
$S(\zt;\gah)$ (up to CDD factors) for the fundamental vector representation.
We argued that in the marginal limit this \uospa\ \zss-matrix reduces to the
exact physical \zss-matrix $S_Y(\zt)$ for the fundamental particles of
the \osp\ current-current model.  We did not prove that the
quantum group symmetry used to determine $S(\zt;\gah)$ and $S_Y(\zt)$ is
exact to all orders in $g$.  
Nevertheless, the fact that we were able to construct a \zss-matrix
satisfying the scattering constraints and having a symmetry algebra agreeing
with the Yangian is strong support for the validity of (\ref{minSYZ}).
As mentioned above, one can try to construct the Yangian charges and in
turn determine $S_Y(\zt)$ from the Yangian symmetry.  However, whereas the
procedure for computing $S$ using quantum group symmetry is, in principle,
well established, this is not the case for the Yangian symmetric situation.
We hope to address this issue for the 
specific model (\ref{seffVZ}) in the future.  Another independent check
of (\ref{minSYZ}) will be to do a thermodynamic Bethe ansatz analysis 
\cite{klassenR,zamotbaR,zzssmtR,ztbarsosR}.  This
is complicated by the fact that $S_Y(\zt)$ is not diagonal.

The $\bh$-dependent \zss-matrix $S(\zt;\gah)$ is itself an interesting
result.  
For $0 < 1/\gah < 1$, $S(\zt;\gah)$ is the \zss-matrix for the supersymmetric
Toda system (\ref{dtacZ}), and reduces to the rational result $S_Y(\zt)$ at
$1/\gah = 0$.  Yet $S(\zt;\gah)$ satisfies all the
scattering constraints for any $1/\gah > 0$, which suggests that it is the 
fundamental ``soliton''
\zss-matrix for some other theory (recall that for $1/\gah > 1$ the Toda
perturbation (\ref{dintVZ}) becomes irrelevant).  In 
the marginal limit this theory
also flows to the \osp\ current-current model.  It is an open question to
determine the theory, in the sense of an 
action, corresponding to (\ref{minSZ}) for
all $\bh$ (or $\gah$).  This theory will be some perturbation of the
$c=0$ CFT (\ref{scftZ}).  Since $S(\zt;\gah)$ has a 
non-trivial pole structure, the
complete \zss-matrix must include the scattering amplitudes for the neutral
and charged bound states.  These amplitudes can be found by applying the
bootstrap principle.


\section*{Acknowledgments}

We would like to thank B.\ Gerganov for useful discussions.  This work is
supported in part by the National Science Foundation, in part through the
National Young Investigator Program.  Z.S.B. also acknowledges support from
the Olin foundation.

\appendix
\setcounter{equation}{0}

\zeqnapp%

\section*{Appendix A}

In this appendix we review the \osp\ superalgebras and give the
relationship between the \ospa\ generators and the 
currents (\ref{holcuZ}),(\ref{aholcuZ}).  Our discussion omits
the Serre relations as they are not needed in this paper.  For the same 
reason, we will also define the affine algebras without the derivation.
Further details on general (affine) Lie superalgebras and quantum 
supergroups, including \osp, may be found 
in \cite{degoliyzR,kacbR,frappatR,sorbaR,cornwellR}.  We will follow most
closely the notations and conventions of \cite{degoliyzR} and \cite{frappatR}
(though unlike \cite{degoliyzR} and \cite{frappatR},
we take a simple root system for \osp\ that is purely fermionic).

\medskip
\noindent
{\it (1) The superalgebras \osp\ and \ospa}

\nopagebreak
\smallskip
The simple Lie superalgebra \osp\ is a $Z_2$-graded algebra with two simple
roots, $\{\al_1,\al_2\}$, and Chevalley generators $\{ e_i, f_i, h_i;i=1,2\}$
satisfying
\zbe
[h_i,h_j] = 0  
\label{AlsanaZ}
\zeqnAalt
\zee
\zadeqn%
\zbe
[h_i,e_j] = a_{ij} e_j,\qquad [h_i,f_j] = - a_{ij} f_j
\zeqnAalt
\zee
\zadeqn%
\zbe
[e_i,f_j] = \delta_{ij} h_i
\zeqnAalt
\zee
\zadeqn%
\zbe
e_i^2 = f_i^2 = 0, \ \ {\rm if\ } a_{ii} = 0.
\zeqnAalt
\zee
\zsetc%
\zeqnapp%
Here (i) $[\cdot\, ,\cdot ]$ denotes the graded Lie bracket
\zbe
[a,b] = a b - (-1)^{d(a) d(b)} b a, \label{AglbZ}
\zee
where $d(x)$ is the parity of $x$: $d(x) = 0$ if $x$ is even or bosonic
and $d(x) = 1$ if $x$ is odd or fermionic (all Cartan generators $h_i$
are even); and (ii) $a_{ij}$ is the
symmetric generalized Cartan matrix defined as
\zbe
a_{ij} = (\al_i,\al_j),  \label{AgscmZ}
\zee
where $(\cdot\, ,\cdot )$ is a fixed invariant bilinear form on the root space.
In contrast with (bosonic) Lie algebras, superalgebras allow several
inequivalent simple root systems.  A common choice is the distinguished
root system, where all simple roots except one are taken to be bosonic.
We will instead work with a purely fermionic \osp\ simple root system
with the parities
\zbe 
d_{1,2} \equiv d(e_{1,2}) = d(f_{1,2}) = 1,\qquad 
\tilde{d}_{1,2} \equiv d(h_{1,2}) = 0.
\zee
For the \osp\ Cartan matrix we take
\zbe
a_{ij} = \lb \begin{array}{cc}
0 & -2 \\
-2 & 0 \\
\end{array} \rb. \label{Ascm1Z}
\zee
A specific realization of the root system is given by
\zbe
\al_1 = (1,i),\qquad \al_2 = (-1,i). \label{ArsZ}
\zee
The Dynkin diagram associated with (\ref{Ascm1Z}) is
\newline
\begin{picture}(468,25)
\put(209,10){\circle{15}}
\put(259,10){\circle{15}}
\put(203.6966,15.3033){\line(1,-1){10.6066}}
\put(203.6966,4.6966){\line(1,1){10.6066}}
\put(253.6966,15.3033){\line(1,-1){10.6066}}
\put(253.6966,4.6966){\line(1,1){10.6066}}
\put(216.0477,12.5651){\line(1,0){35.9046}}
\put(216.0477,7.4348){\line(1,0){35.9046}}
\end{picture}

The untwisted affine Lie superalgebra \ospa\ contains an additional root
$\al_0 = - \psi$, where $\psi$ is the highest root of \osp.  Explicitly we 
have
\zbe
\al_0 = -(\al_1 + \al_2) = (0,-2i).  \label{AasrZ}
\zee
This additional root is necessarily even, with $d_0 \equiv d(e_0)=d(f_0)=0$.
The affine generators $\{e_i,f_i,h_i; i=0,1,2\}$ satisfy the same relations
(\ref{AlsanaZ}) as for \osp.
(If we included the derivation, $\al_0$ would be given by $\delta-\psi$,
where $\delta$ is the minimal imaginary root of \ospa.  However, since
$\delta$ satisfies $(\delta,\delta) = (\delta,\al_{1,2})=0$, the defining
relations (\ref{AlsanaZ}) are unchanged.)
The affine Cartan matrix is 
\zbe 
a_{ij} = \lb
\begin{array}{ccc}
-4 & 2 & 2 \\
2 & 0 & -2 \\
2 & -2 & 0 \\
\end{array} \rb.
\label{Ascm2Z}
\zee

\noindent
Note that in section 3 we find $h_0 = -(h_1 + h_2)$, thus there $h_0$ is not
an independent generator.  This implies that the central extension is zero.
For a non-zero central extension all the Cartan generators $\{h_0,h_1,h_2\}$
satisfying (\ref{AlsanaZ}) are independent.

\medskip
\noindent
{\it (2) The quantum superalgebras \uosp\ and \uospa}

\smallskip
The quantum superalgebras (or supergroups) \uosp\ and \uospa\ are deformations
of the universal enveloping algebras for \osp\ and \ospa.  As such they
are (unital) $Z_2$-graded associative algebras generated by
$\{e_i,f_i,h_i\}$, where $i=1,2$ for \uosp\ and $i=0,1,2$ for \uospa, modulo
the relations
\zbe
[h_i,h_j] = 0  
\label{AlsaaZ}
\zeqnAalt
\zee
\zadeqn%
\zbe
[h_i,e_j] = a_{ij} e_j,\qquad [h_i,f_j] = - a_{ij} f_j
\zeqnAalt
\zee
\zadeqn%
\zbe
[e_i,f_j] = \delta_{ij} \frac{ q^{h_i} - q^{-h_i}}{q - q^{-1}}
\zeqnAalt
\zee
\zadeqn%
\zbe
e_i^2 = f_i^2 = 0, \ \ {\rm if\ } a_{ii} = 0.
\zeqnAalt
\zee
\zsetc%
\zeqnapp%
Here $q$ is an arbitrary non-zero complex number.  In the limit 
$q\rightarrow 1$, (\ref{AlsaaZ}) reduces to (\ref{AlsanaZ}).  The 
expressions $q^{\pm h_i}$ are
understood as infinite power series in $h_i$.  We can alternatively write
(\ref{AlsaaZ}a) and (\ref{AlsaaZ}b) as
\zbe
q^h q^{h^\prime} = q^{h^\prime} q^h, \quad h,h^\prime \in \{\pm h_i\}
\label{AlsaabZ}
\zeqnAalt
\zee
\zadeqn%
\zbe
q^{h_i} q^{-h_i} = q^{-h_i} q^{h_i} = 1
\zeqnAalt
\zee
\zadeqn%
\zbe
q^{h_i} e_j q^{-h_i} = q^{a_{ij}} e_j, \qquad
q^{h_i} f_j q^{-h_i} = q^{-a_{ij}} f_j.
\zeqnAalt
\zee
\zsetc%
\zeqnapp%
These quantum algebras can be endowed with a Hopf algebraic structure.
The comultiplication $\zD$ is defined as
\zbe
\zD(h_i) = h_i\otimes 1 + 1\otimes h_i
\label{AlsacomZ}
\zeqnAalt
\zee
\zadeqn%
\zbe
\zD(e_i) = e_i \otimes q^{-h_i/2} + q^{h_i/2} \otimes e_i
\zeqnAalt
\zee
\zadeqn%
\zbe
\zD(f_i) = f_i \otimes q^{-h_i/2} + q^{h_i/2} \otimes f_i,
\zeqnAalt
\zee
\zsetc%
\zeqnapp%
or equivalently in place of (\ref{AlsacomZ}a)
\zbe
\zD(q^{\pm h_i}) = q^{\pm h_i} \otimes q^{\pm h_i}.
\label{AlsacomhbZ}
\zee
The antipode $\cs$ and counit $\varepsilon$ are
\[
\cs(h_i) = - h_i
\]
\[
\cs(e_i) = - q^{-(\al_i,\al_i)/2} e_i
\]
\zbe
\cs(f_i) = - q^{(\al_i,\al_i)/2} f_i \label{AlsaapZ}
\zee
\zbe
\varepsilon(h_i) = \varepsilon(e_i)=\varepsilon(f_i) = 0,
\qquad \varepsilon(1) = 1. 
\zee

A thorough discussion of the representation theory for the \osp\ 
algebras can be found in \cite{snrR,zhgoR,zhangR,ligoR,kacaR,zhgobrybR}.

\medskip
\noindent
{\it (3) The \ospa\ currents}

The super-currents (\ref{holcuZ}) have Laurent 
expansions of the form (concentrating only on the holomorphic sector)
\zbe
\zo(z) = \sum_{n} z^{-n-1} \zo_n. \label{AscleZ}
\zee
The algebra satisfied by the modes $\{ \zo_n \}$ can be obtained using
the OPE's (\ref{holopeZ}) and the formula
\zbe
[A,B] = \oint_0 d\omega\,\oint_\omega dz\, a(z) b(\omega),
\label{AcopeZ}
\zee
for
\zbe
A = \oint_0 dz\, a(z),\qquad B = \oint_0 dz\, b(z).
\label{AopdefZ}
\zee
For the modes $\zo_m$ and $\zo_n^\prime$, the (graded) commutator is
\zbe
[\zo_m,\zo_n^\prime ] = \oint_0 d\omega\, \omega^n\oint_\omega dz\,
z^m \zo(z) \zo^\prime(\omega).
\label{AmcopeZ}
\zee
To prove the equivalence between the current algebra and the level one
\ospa\ algebra, we need to show that (\ref{AmcopeZ}) agrees 
with (\ref{AlsanaZ}).  This amounts to
correctly identifying the modes with the generators
$\{e_i,f_i,h_i;i=0,1,2\}$.  We find the following relations (here
$(\zo)_m \equiv \zo_m$)
\[
e_0 = \fb \kh_{-1},\qquad f_0 = \fb K_{+1},\qquad h_0 = 1 + 2 H_0
\]
\[
e_1 = i (G_-)_0,\qquad f_1 = i (\gh_+)_0,\qquad h_1 = -(H_0 + J_0)
\]
\zbe
e_2 = i (G_+)_0,\qquad f_1 = i (\gh_-)_0,\qquad h_2 = -(H_0 - J_0).
\label{AmgrelZ}
\zee
Note that $h_0 \neq - (h_1+h_2)$, but rather  $h_0 = k - (h_1 + h_2)$
where $k=1$ is the level.  This differs from the quantum group 
charges (\ref{erel2Z})
where the level, or central extension, is zero and $T^0=-(T^1 + T^2)$.
One can check that (\ref{AmgrelZ}) is consistent 
with (\ref{holopeZ}) and (\ref{AlsanaZ}).

In the Cartan-Weyl basis, the conserved currents for the conformal field
theory (\ref{scftZ}) are
\[
{\cal{J}}(z) = e_0 \fb \kh + e_1 (i G_-) + e_2 (i G_+) +
f_0\fb K + f_1 (i \gh_-) + f_2 (i\gh_+)
+ \fa (h_2 - h_1) J - \fa (h_2 + h_1) H
\]
\zbe
{\overline{\cal{J}}}(\zb) = e_0 \fb \khb + e_1 (-i \gb_-) + e_2 (-i \gb_+) +
f_0\fb \kb + f_1 (-i \ghb_+) + f_2 (-i\ghb_-)
+ \fa (h_2 - h_1) \jb - \fa (h_2 + h_1) \hb.
\label{AfulljZ}
\zee
A current-current interaction $\Phi_{\rm int}^{\rm cc}$ is given by
\zbe
\Phi_{\rm int}^{\rm cc} \propto
{\rm str}\lb \overline{{\cal J}}{\cal J} \rb,
\label{AstrjjZ}
\zee
where the constant of proportionality depends on the particular
representation.  Evaluating (\ref{AstrjjZ}) using the 
representation matrices (\ref{chrepZ}) and (\ref{afchZ}) we get
\zbe
{\rm str}\lb \overline{{\cal J}}{\cal J}\rb =
-2 \left[ \jb J - \hb H + \fa(\kb \kh + \khb K) + \gb_- \gh_+
- \ghb_- G_+ + \gb_+ \gh_- - \ghb_+ G_- \right].
\zee
This is identical to (\ref{intVcuZ}) hence confirming that $\Phi_V$ is indeed a
current-current perturbation.


\newcommand{\Rzf}{\textbf}

\newcommand\adpR{Adv.\ Phys.\ }
\newcommand\aspmR{Adv. Studies Pure Math.\ }
\newcommand\anpR{Ann.\ Phys.\ } 
\newcommand\cmpR{Commun.\ Math.\ Phys.\ }
\newcommand\ijmpaR{Int.\ J.\ Mod.\ Phys.\ A\ }
\newcommand\ijmpbR{Int.\ J.\ Mod.\ Phys.\ B\ }
\newcommand\jmpR{J.\ Math.\ Phys.\ }
\newcommand\jpaR{J.\ Phys.\ A\ } 
\newcommand\jpcR{J.\ Phys.\ C\ }
\newcommand\jsmR{J.\ Sov.\ Math.\ }
\newcommand\jspR{J.\ Stat.\ Phys.\ }
\newcommand\lmpR{Lett.\ Math.\ Phys.\ }
\newcommand\mplaR{Mod.\ Phys.\ Lett.\ A\ }
\newcommand\npbR{Nucl.\ Phys.\ B\ }
\newcommand\prR{Phys.\ Rev.\ }
\newcommand\praR{Phys.\ Rev.\ A\ }
\newcommand\prbR{Phys.\ Rev.\ B\ }
\newcommand\prdR{Phys.\ Rev.\ D\ }
\newcommand\plR{Phys.\ Lett.\ }
\newcommand\plaR{Phys.\ Lett.\ A\ }
\newcommand\plbR{Phys.\ Lett.\ B\ }
\newcommand\prlR{Phys.\ Rev.\ Lett.\ }
\newcommand\ptpR{Prog.\ Theor.\ Phys.\ }
\newcommand\rmpR{Rev.\ Mod.\ Phys.\ }
\newcommand\sjnpR{Sov.\ J.\ Nucl.\ Phys.\ }
\newcommand\smdR{Sov.\ Math.\ Dokl.\ }
\newcommand\spjR{Sov.\ Phys.\ JETP [Zh.\ Eksp.\ Teor.\ Fiz.]\ }
\newcommand\spjlR{JETP Lett.\ }
\newcommand\zpR{Z.\ Phys.\ }



\end{document}